\documentclass[12pt, final,onecolumn,journal]{IEEEtran}
\usepackage{cite,acronym,paralist,xspace}
\usepackage[cmex10]{amsmath}
\usepackage{amssymb,amsfonts,amsthm}
\usepackage{algorithmic}
\usepackage{graphicx}
\usepackage{pgf}
\usepackage{textcomp}
\usepackage{xcolor}
\usepackage{stmaryrd} 
\usepackage{dsfont}
\usepackage{balance}
\usepackage{diagbox}
\usepackage{enumitem}
\usepackage{cite}
\usepackage{placeins}

\acrodef{ACDIS}[ACDIS]{Adaptive Communication Decision and Information Systems}
\acrodef{AEP}{Asymptotic Equipartition Property}
\acrodef{AoA}{Angle of Arrival}
\acrodef{AWGN}{Additive White Gaussian Noise}
\acrodef{AVC}[AVC]{Arbitrarily Varying Channel}
\acrodef{BER}{Bit-Error-Rate}
\acrodef{BEC}{Binary Erasure Channel}
\acrodef{BPSK}{Binary Phase-Shift Keying}
\acrodef{BSC}{Binary Symmetric Channel}
\acrodef{BICM}[BICM]{Bit-Interleaved Coded-Modulation}
\acrodef{CDF}[CDF]{Cumulative Distribution Function}
\acrodef{CGF}[CGF]{Cumulant Generating Function}
\acrodef{CLT}[CLT]{Central Limit Theorem}
\acrodef{cq}[c-q]{Classical-Quantum}
\acrodef{CSI}[CSI]{Channel State Information}
\acrodef{DMC}[DMC]{Discrete Memoryless Channel}
\acrodef{DMS}[DMS]{Discrete Memoryless Source}
\acrodef{ERM}[ERM]{Empirical Risk Minimization}
\acrodef{FER}[FER]{Frame Error Rate}
\acrodef{ICA}[ICA]{Independent Component Analysis}
\acrodef{iid}[i.i.d.]{independent and identically distributed}
\acrodef{IoT}[IoT]{Internet of Things}
\acrodef{KKT}[KKT]{Karush-Kuhn Tucker}
\acrodef{LASSO}[LASSO]{Least Absolute Shrinkage and Selection Operator}
\acrodef{LPD}[LPD]{Low Probability of Detection}
\acrodef{LDPC}[LDPC]{Low-Density Parity-Check}
\acrodef{LLMS}[LLMS]{Linear Least Mean Square}
\acrodef{LMS}[LMS]{Least Mean Square}
\acrodef{MAC}[MAC]{multiple-access channel}
\acrodef{MGF}[MGF]{Moment Generating Function}
\acrodef{MLC}[MLC]{Multi-Level Coding}
\acrodef{MLE}[MLE]{Maximum Likelihood Estimate}
\acrodef{MIMO}[MIMO]{Multiple-Input Multiple-Output}
\acrodef{MISO}{Multiple-Input Single-Output}
\acrodef{MSD}[MSD]{Multi-Stage Decoding}
\acrodef{MMSE}[MMSE]{Minimum Mean-Square Error}
\acrodef{PAC}[PAC]{Probably Approximately Correct}
\acrodef{PCA}[PCA]{Principal Component Analysis}
\acrodef{PDF}[PDF]{Probability Density Function}
\acrodef{PMF}[PMF]{Probability Mass Function}
\acrodef{POVM}[POVM]{Positive Operator-Valued Measure}
\acrodef{PVM}[PVM]{Projection-Valued Measure}
\acrodef{PPM}[PPM]{Pulse Position Modulation}
\acrodef{PSD}{Power Spectral Density}
\acrodef{PSK}{Phase Shift Keying}
\acrodef{QKD}{Quantum Key Distribution}
\acrodef{ROC}{Receiver Operating Characteristic}
\acrodef{CVQKD}{Continuous-Variable \ac{QKD}}
\acrodef{QPSK}{Quadrature Phase-Shift Keying}
\acrodef{RV}{random variable}
\acrodef{SIMO}{Single-Input Multiple-Output}
\acrodef{SNR}{Signal-to-Noise Ratio}
\acrodef{SVM}[SVM]{Support Vector Machine}
\acrodef{TPCP}{Trace-Preserving Completely-Positive}
\acrodef{wrt}[w.r.t.]{with respect to}
\acrodef{WSS}{Wide Sense Stationary}

\newcommand{\calA}{\mathcal{A}}

\newcommand{\calC}{\mathcal{C}}

\newcommand{\calM}{\mathcal{M}}

\newcommand{\calN}{\mathcal{N}}
\newcommand{\bbN}{\mathbb{N}}

\newcommand{\calO}{\mathcal{O}}

\newcommand{\bbR}{\mathbb{R}}

\newcommand{\calS}{\mathcal{S}}

\newcommand{\calU}{\mathcal{U}}

\newcommand{\bfX}{\mathbf{X}}
\newcommand{\calX}{\mathcal{X}}

\newcommand{\bfx}{\mathbf{x}}

\newcommand{\bfY}{\mathbf{Y}}
\newcommand{\calY}{\mathcal{Y}}

\newcommand{\bfy}{\mathbf{y}}

\newcommand{\bfZ}{\mathbf{Z}}
\newcommand{\calZ}{\mathcal{Z}}

\newcommand{\bfz}{\mathbf{z}}

\newcommand{\card}[1]{{|{#1}|}}
\newcommand{\abs}[1]{{\left|{#1}\right|}}

\newcommand{\norm}[2][]{{\left\Vert{#2}\right\Vert}_{#1}}

\newcommand{\set}[1]{{\{#1\}}}

\newcommand{\intseq}[2]{[{#1};{#2}]}
\newcommand{\eqdef}{\triangleq}
\newcommand{\E}[2][]{{\mathbb{E}_{#1}\left[#2\right]}}
\newcommand{\Var}[2][]{{\text{Var}_{#1}\left(#2\right)}}

\renewcommand{\P}[2][]{{\mathbb{P}_{#1}\left(#2\right)}}
\newcommand{\indic}[1]{{\mathbf{1}\{#1\}}}

\newcommand{\D}[3][]{{\mathbb{D}_{#1}}\!\left(#2\,\middle\Vert\,#3\right)} 
\newcommand{\V}[3][]{{\mathbb{V}_{#1}}\!\left(#2,#3\right)} 
\newcommand{\avgI}[1]{{{\mathbb{I}}\!\left(#1\right)}} 
\newcommand{\avgH}[1]{{\mathbb{H}}\!\left(#1\right)}
\newcommand{\h}[1]{{{h}}\left(#1\right)} 
\newcommand{\Hb}[1]{{h_b}\left(#1\right)} 
\newcommand{\chidist}[3]{{\chi}_{#1}\!\left(#2\,\middle\Vert\,#3\right)} 

\newcommand{\wt}[1]{\ensuremath{\mbox{wt}(#1)}} 

\renewcommand{\leq}{\leqslant} 
\renewcommand{\geq}{\geqslant} 

\newcommand{\limn}[2][n\rightarrow\infty]{\underset{#1}{\mbox{\textnormal{lim}}\;}{#2}}
\newcommand{\Ww}{W_{Z|X}}
\newcommand{\Wwn}{W_{\bfZ|\bfX}^{\otimes n}}

\newcommand{\Wb}{W_{Y|X}}
\newcommand{\Wbn}{W_{\bfY|\bfX}^{\otimes n}}
\newcommand{\WbN}{W_{\bfY|\bfX}^{\otimes N}}
\newcommand{\Pinn}{P_{0}}
\newcommand{\Pinf}{P_{1}}
\newcommand{\Qinn}{Q_{0}}
\newcommand{\Qinf}{Q_{1}}
\newcommand{\Qcoden}{\widehat{Q}_Z^n}
\newcommand{\QcodeN}{\widehat{Q}_Z^{N}}

\usepackage[nameinlink]{cleveref}

\newcounter{numrellocal}
\renewcommand{\thenumrellocal}{\alph{numrellocal}}
\newcounter{numrelglobal}

\makeatletter
\newcommand{\labrel}[2]{
  \stepcounter{numrellocal}
  \refstepcounter{numrelglobal}
  \ltx@label[equation]{#2}
  \stackrel{\textnormal{(\thenumrellocal)}}{\mathstrut{#1}}
}
\makeatother

\everydisplay\expandafter{\the\everydisplay\setcounter{numrellocal}{0}} 

\allowdisplaybreaks
\newtheorem{theorem}{Theorem}%
\newtheorem{proposition}[theorem]{Proposition}%
\newtheorem{lemma}[theorem]{Lemma}%
\newtheorem{definition}[theorem]{Definition}%
\newtheorem{remark}{Remark}
\Crefname{lemma}{Lemma}{Lemmas}
\crefname{lemma}{lemma}{lemmas}

\begin{document}
\author{Shi-Yuan Wang, Keerthi S. K. Arumugam, Matthieu R. Bloch \thanks{This work was supported by the National Science Foundation under Award 1955401. The results presented here build upon preliminary results presented at the 2016 IEEE Information Theory Workshop~\cite{Arumugam2016a}.}
  \thanks{This works has been accepted to IEEE Transactions on Information Theory. The published version by IEEE will be associated with DOI: 10.1109/TIT.2025.3589575.}
\thanks{\copyright 2025 IEEE.  Personal use of this material is permitted.  Permission from IEEE must be obtained for all other uses, in any current or future media, including reprinting/republishing this material for advertising or promotional purposes, creating new collective works, for resale or redistribution to servers or lists, or reuse of any copyrighted component of this work in other works.}}
\title{Bounds on Covert Capacity with \\Sub-Exponential Random Slot Selection}
\maketitle

\allowbreak
\begin{abstract}
  We consider the problem of covert communication with random slot selection over binary-input \acp{DMC} and \ac{AWGN} channels, in which a transmitter attempts to reliably communicate with a legitimate receiver while simultaneously maintaining covertness \ac{wrt} an eavesdropper. Covertness refers to the inability of the eavesdropper to distinguish the transmission of a message from the absence of communication, modeled by the transmission of a fixed channel input. Random slot selection refers to the transmitter's ability to send a codeword in a time slot with known boundaries selected uniformly at random among a predetermined number of slots. Our main contribution is to develop bounds for the information-theoretic limit of communication in this model, called the covert capacity, when the number of time slots scales sub-exponentially with the codeword length. Our upper and lower bounds for the covert capacity are within a multiplicative factor of $\sqrt{2}$ independent of the channel. This result partially fills a characterization gap between the covert capacity without random slot selection and the covert capacity with random selection among an exponential number of slots in the codeword length. Our key technical contributions consist of
  \begin{inparaenum}[i)]
  \item  a tight upper bound for the relative entropy characterizing the effect of random slot selection on the covertness constraint in our achievability proof;
  \item  a careful converse analysis to characterize the maximum allowable weight or power of codewords to meet the covertness constraint.
  \end{inparaenum}
  Our results suggest that, unlike the case without random slot selection, the choice of covertness metric does not change the covert capacity in the presence of random slot selection. 
\end{abstract}

\section{Introduction}
\label{sec:intro}

Covert communications, also known as communications with low-probability of detection, aim at hiding the presence of communication signals from eavesdroppers. The information-theoretic study of covert communications has attracted renewed interest following the characterization of a square-root-law for covert communications~\cite{Bash2013} and the characterization of the associated covert capacity~\cite{Wang2016b,Bloch2016}. Specifically, information-theoretic covertness refers to an eavesdropper's inability to distinguish the presence or absence of transmissions from his observations, measured in terms of the distance between the observation statistics induced by transmissions and those induced by a fixed channel input symbol representing the absence of transmission. The square-root law shows that, for \ac{AWGN} channels and all but a few \acp{DMC}~\cite{Wang2016b,Bloch2016,bouetteCovertCommunicationTwo2023}, the number of bits that can be transmitted reliably while avoiding detection by an eavesdropper over $n$ channel uses, must asymptotically scale no faster than the square-root of the blocklength $O(\sqrt{n})$. The covert capacity characterizes the optimal constant hidden behind the $O(\sqrt{n})$ notation. We note that the notion of low-probability of interception is a different problem, captured through the information-theoretic notion of stealth~\cite{Hou2014,Lin2020Stealthy,Lentner2020}. Unlike covertness, stealth is measured in terms of the distance between the observation statistics induced by transmissions and those induced by a random but known distribution of channel input symbols representing a typical behavior to mimic. Stealth communications are typically not subject to the square-root law~\cite{Lentner2020} and naturally combine with secrecy through the notion of \emph{effective secrecy}, by which both the presence of a message and the content of a message can be simultaneously hidden from an eavesdropper~\cite{Hou2014,Hou2017}. 

The realization that covertness fundamentally restricts communication to happen at zero rate has naturally prompted the investigation of models for which the square-root law could be circumvented. For \ac{MIMO} \ac{AWGN} channels, the number of reliable and covert bits can scale as $O(n)$ when it is possible to transmit information in the null space of the eavesdropper~\cite{Wang2021,Bendary2021}. For classical-quantum channels, the existence of shared entanglement hidden from the adversary allows the number of reliable and covert bits to scale as $O(\sqrt{n}\log n)$~\cite{Gagatsos2020Covert,Zlotnick2023Entanglement,Wang2024Resource}. For \ac{AWGN} channels and \acp{DMC}, the existence of eavesdropper's uncertainty also breaks the square-root law. Several types of uncertainties have been considered:
\begin{inparaenum}[i)]
\item uncertainty in the channel noise allows the number of covert and reliable bits to scale as $O(n)$, either through jamming~\cite{Sobers2017}, asymmetry in channel knowledge~\cite{Lee2018a,ZivariFard2020}, or inherent uncertainty~\cite{Che2014a};
\item uncertainty in which users are communicating in a multi-user random access model~\cite{Hayashi2024Covert};
\item uncertainty in the timing of transmission, hereafter referred to as \emph{asynchronism}~\cite{Tchamkerten2009}, allows the number of covert and reliable bits to scale everywhere in between $O(\sqrt{n})$ to $O(n)$ depending on the level of randomization in the timing of transmission~\cite{Bash2016,Goeckel2016,Arumugam2016a,Dani2021Covert}.
\end{inparaenum}
The analysis of information-theoretic limits with unusual scalings with the codeword length presents additional challenges compared to the traditional notion of channel capacity: one now needs to first identify the optimal scaling and then the optimal constant in front of the scaling. While perhaps uncommon, such scaling issues also appear in other areas of information theory, such as those related to the notion of identification capacity~\cite{ibrahimIdentificationEffectiveSecrecy2021,rosenbergerCapacityBoundsIdentification2023}.

\begin{figure}[t]
  \centering
  \includegraphics[width=0.8\textwidth]{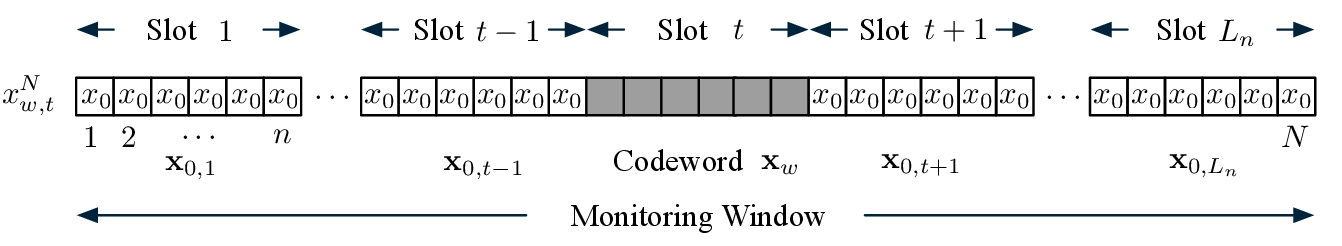}
  \caption{Illustration of random slot selection for the frame synchronous model considered here. The transmitter, encoding a message $w$ into a codeword $\mathbf{x}_w$ of length $n$, places $\bfx_w$ in a slot $t$ chosen uniformly at random out of $L_n$ consecutive slots. The other slots are left unused, modeled by the use of a fixed symbol $x_0$ representing the absence of communication.}
  \label{fig:codeword}
\end{figure}

The present work studies the information-theoretic limits of covert communications in the presence of transmission timing uncertainty by considering a frame synchronous model in which the transmitter can randomize the placement of his codeword among a set of frames as in~\cite{Bash2016,Goeckel2016}. Specifically, as illustrated in Fig.~\ref{fig:codeword}, the transmitter sends a message $w$ by first encoding it into a codeword $\mathbf{x}_w$ of length $n$ and then placing it in a slot $t$ chosen uniformly at random out of $L_n$ consecutive slots. The case $L_n=1$ recovers the covert communication model subject to the square-root law. The case where $L_n$ is exponential in the codeword length $n$ corresponds to the situation called ``exponential asynchronism'', in which case \cite[Theorem 1]{Dani2021Covert} shows that the number of reliable and covert bits may scale as $O(n)$ and characterizes the covert capacity for exponential asynchronous \acp{BSC}.\footnote{The terminology ``exponential asynchronism'' adopted by \cite{Dani2021Covert} is indeed random slot selection and is therefore different from the strong asynchronous regime of~\cite{Tchamkerten2009}.} The case where $L_n$ is sub-exponential has been studied in~\cite{Bash2016, Goeckel2016, Arumugam2016a, Dani2021Covert} but all these prior works fall short of characterizing the covert capacity. The present work partially fills that gap, by characterizing the covert capacity with sub-exponential random slot selection within a multiplicative $\sqrt{2}$ factor. The techniques employed to establish our results carefully extend ones that already proved successful to study the covert capacity~\cite{Bloch2016,Arumugam2016a,Tahmasbi2019}. We conjecture that our upper bound is the correct covert capacity although we were unable to close the gap.

The remaining of the paper is organized as follows. In Section~\ref{sec:notation}, we set the notation used throughout the paper. In Section~\ref{sec:model-and-results}, we introduce the model of covert communications along with formal definitions of different levels of random slot selection. We also introduce our characterizations of the covert capacity for \acp{DMC} and \ac{AWGN} channels under sub-exponential random slot selection. We relegate all proofs to Section~\ref{sec:proofs} and Section~\ref{sec:proof-thm-Gaussian} to streamline the presentation. In Section~\ref{sec:concluding-remark}, we conclude with discussion of possible extensions and associated challenges.

\section{Notation}
\label{sec:notation}
Let $\bbR_+$ and $\bbN_*$ denote all non-negative real numbers and all positive integers, respectively.
For any set $\Omega$, the indicator function is defined as $\indic{\omega\in\Omega} = 1$ if $\omega\in \Omega$ and $0$ otherwise.
For any discrete set $\calX$ and $n\in\bbN_*$, a sequence of length $n$ is implicitly denoted $\mathbf{x}\eqdef(x_1,\cdots,x_n)\in \calX^n$, while $x^i\eqdef (x_1,\cdots,x_i)\in \calX^i$ denotes a sequence of length $i$. We also let $\bfx_{\intseq{i}{j}}$, $j>i>0$ denote the sequence $(x_i,x_{i+1},\cdots, x_{j-1}, x_{j})$.

$\avgH{\cdot}$, $\h{\cdot}$, $\avgI{\cdot;\cdot}$, and $\Hb{\cdot}$ denote the usual entropy, differential entropy, mutual information, and binary entropy function, respectively.

\sloppy For two distributions $P_X$ and $Q_X$ over the same set $\calX$,
the relative entropy is $\D{P_X}{Q_X}\eqdef\sum_{x}P_X(x)\log\frac{P_X(x)}{Q_X(x)}$, the variational (total variation) distance is $\V{P_X}{Q_X}\eqdef\frac{1}{2}\sum_x\abs{P_X(x)-Q_X(x)}$, and the chi-squared distance is $\chidist{2}{P_X}{Q_X}\eqdef\sum_x\frac{\left(P_X(x)-Q_X(x)\right)^2}{Q_X(x)}$. We let $P_X\ll Q_X$ denote the absolute continuity \ac{wrt} $Q_X$, i.e., if for any $x\in\calX$, $Q_X(x)=0$ implies $P_X(x)=0$.

For a continuous alphabet $\calX=\bbR$ and any two distributions $P,~Q$ with densities $f_P,~f_Q$, respectively, the variational (total variation) distance between $P$ and $Q$ is defined as
$\V{P}{Q}\eqdef\frac{1}{2}\int_{\bbR}|f_P\left(x\right)-f_Q\left(x\right)|dx$ or equivalently 
$\V{P_X}{Q_X}=\sup_{\calS\subseteq\bbR}|P\left(\calS\right)-Q\left(\calS\right)|$.

The relative entropy between $P_X$ and $Q_X$ is defined as 
$\D{P_X}{Q_X}\eqdef\int_{\bbR}f_P(x)\times\log\frac{f_P(x)}{f_Q(x)}dx$.
Pinsker's inequality ensures that $$\V{P_X}{Q_X}^2\leq\frac{1}{2}\min(\D{P_X}{Q_X},\D{Q_X}{P_X}).$$ Let $X\in\calX$ and $Y\in\calY$ be jointly distributed random variables acccording to $P\cdot W$ where $W:(x, y)\mapsto W(y|x)$ is a transition probability from $\calX\to\calY$. We define the marginal distribution of $Y$ as $P\circ W$.

For $a,b\in\bbR$ such that $\lfloor a\rfloor\leq \lceil b\rceil$, we define $\intseq{a}{b}\eqdef\{\lfloor a\rfloor, \lfloor a\rfloor+1, \cdots, \lceil b\rceil-1, \lceil b\rceil\}$; otherwise $\intseq{a}{b}\eqdef\emptyset$. In addition, for any $x\in\bbR$, we let $\abs{x}^+$ denote $\max(x,0)$. For any $x\in\bbR$, we define the $Q$-function $Q(x)\eqdef\int_x^\infty\frac{1}{\sqrt{2\pi}}e^{\frac{-x^2}{2}}dx$ and its inverse function $Q^{-1}(\cdot)$. For $\calX=\{x_0, x_1\}$, we also define $\wt{\bfx}\eqdef\sum_{i=1}^n\indic{x_i=x_1}$. Finally, throughout the paper, $\log$ is \ac{wrt} base $e$, and therefore all the information quantities should be understood in \emph{nats}.

\section{Model and Results}
\label{sec:model-and-results}

\begin{figure}[t]
  \centering
  \includegraphics[width=0.8\textwidth]{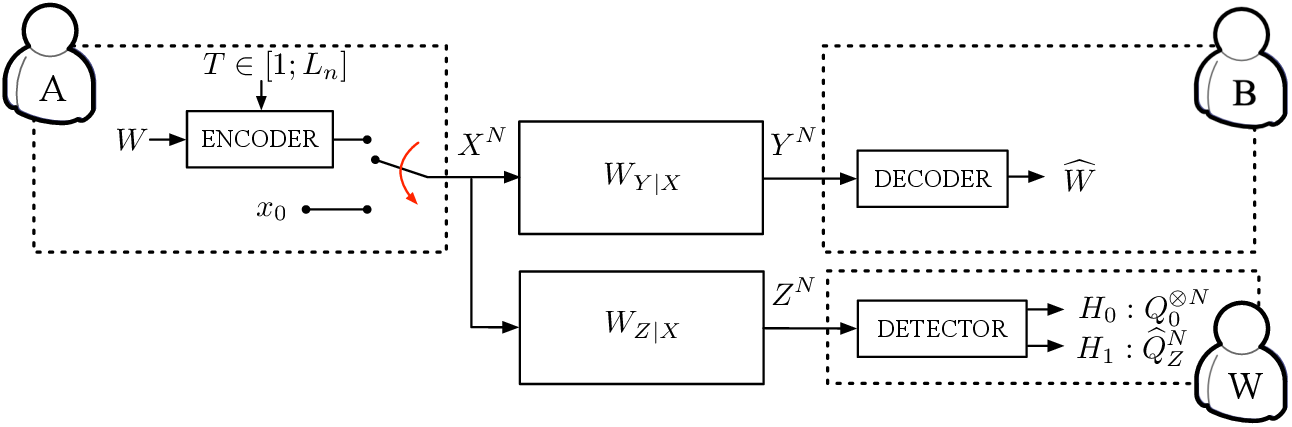}
  \caption{Model of covert communication with random slot selection.}
  \label{fig:model}
\end{figure}

We consider the model of covert communication with random slot selection illustrated in Fig.~\ref{fig:model}, in which a legitimate transmitter, Alice, attempts to communicate with a legitimate receiver, Bob, while avoiding the detection of an adversary, Willie. Specifically, the communication channel is modeled as a memoryless channel with transition probability $W_{YZ|X}$, in which the input symbols in $\calX$ are controlled by Alice while the output symbols in $\calY$ and $\calZ$ are observed by Bob and Willie, respectively. We shall consider the case of binary-input \acp{DMC}, in which case we assume that $\calX\eqdef\set{x_0,x_1}$ where $x_0$ models the absence of communication, $\card{\calY}<\infty$, $\card{\calZ}<\infty$. We set for $j\in\set{0,1}$ $P_j\eqdef W_{Y|X=x_j}$ and $Q_j\eqdef W_{Z|X=x_j}$ and assume that $\Qinf\ll \Qinn$, $\Pinf\ll \Pinn$, and $\Qinf\neq\Qinn$ to avoid creating an unfair advantage in the channel statistics for one of the parties as discussed in~\cite[Appendix G]{Bloch2016}. We shall also consider \ac{AWGN} channels, in which case $\calX=\calY=\calZ=\bbR$ and the input-output relationships of the channel are given at each time instant $i$ by
\begin{align}\label{eq:AWGN-model}
  Y_i=  X_i + N_{b,i}\qquad Z_i=X_i+N_{w,i},
\end{align}
where $\set{N_{b,i}}$ and $\set{N_{w,i}}$ are \ac{iid} Gaussian noises with zero mean and variance $\sigma_b^2$ and $\sigma_w^2$, respectively. We set $x_0=0$ in that case.
{To avoid the need for pre-shared keys between Alice and Bob, we further restrict our analysis to \emph{keyless} schemes and assume $\D{P_1}{P_0}>\D{Q_1}{Q_0}$ for the \ac{DMC} scenario and $\sigma_w^2>\sigma_b^2$ for the \ac{AWGN} scenario.}

Alice encodes a uniformly distributed message $W\in\intseq{1}{M}$ into a codeword $\bfX_w\in\calX^n$ of blocklength $n$ using a publicly known encoding function. {Alice may place} the codeword $\bfX_w$ in one of $L_n\in\bbN_*$ slots whose boundaries are known to all parties, so that the overall blocklength for transmission is $N\eqdef n\cdot L_n$. We assume that the slot $T$ used for transmission is chosen uniformly at random by the encoder, i.e., $T\sim\calU(1, L_n)$, and $T$ is unknown to both Bob and Willie. If $T=t$ and $W=w$, we use $x^N_{w, t}$ to represent the codeword $\bfx_w$ transmitted in the $t$-th slot, i.e., $x_{w,t}^N\eqdef(\bfx_{0, 1}, \cdots, \bfx_{0, t-1}, {\bfx}_w, \bfx_{0, t+1}, \cdots, \bfx_{0, L_n})$, as shown in Fig.~\ref{fig:codeword}, where $\bfx_{0,i}$ represents $x_0^n$ in the $i$-th slot. The case $L_n=1$ corresponds to the absence of random slot selection. The case $L_n=\Omega(e^{A n})$ for some $A>0$ corresponds to exponential random slot selection studied in~\cite{Dani2021Covert}. The case $L_n=\omega(1)\cap \bigcap_{c>0} o(c^n)$ corresponds to the \emph{sub-exponential} random slot selection and is the one of interest here.

Upon receiving a noisy sequence $Y^{N}$, Bob attempts to form an estimate $\widehat{W}\eqdef g(Y^{N})$ of the message transmitted by Alice. The reliability performance is measured using the average probability of error $P_e^{(n)}\eqdef \P{\smash{\widehat{W}\neq W}}$. Simultaneously, Alice attempts to escape detection by Willie who observes the noisy sequence $Z^{N}$. The covertness performance is measured by how much the distribution induced by the coding scheme $\QcodeN$ differs from the reference distribution $Q_0^{\otimes N}$, induced when Alice does not communicate and only sets the input to $x_0^{N}$. Specifically, the distribution induced at Willie's channel output by the coding scheme is $\QcodeN(z^N)\eqdef\E[T]{\smash{\widehat{Q}_{Z, T}^N(z^N)}}$, where
\begin{align}
  \label{eq:Q-N-T-hat}
  \widehat{Q}_{Z,t}^N(z^N)\eqdef&\left(\prod_{\ell=1}^{t-1}\Qinn^{\otimes n}(\bfz_{\intseq{n(\ell-1)+1}{n\ell}})\right)\nonumber\\
  &\times\left(\frac{1}{M}\sum_{w=1}^M\Wwn(\bfz_{\intseq{n(t-1)+1}{nt}}|{\bfx}_w)\right)\left(\prod_{\ell=t+1}^{L_n}\Qinn^{\otimes n}(\bfz_{\intseq{n(\ell-1)+1}{n\ell}})\right).
\end{align}
\sloppy We measure covertness using the variational distance (total variation distance) $V^{(n)}\eqdef \V{\smash{\widehat{Q}^N_{Z}}}{Q_0^{\otimes N}}$. The goal of Willie's hypothesis testing is to distinguish between the null hypothesis $H_0:Z^N\sim\Qinn^{\otimes n}$ and the alternative hypothesis $H_1:Z^N\sim\QcodeN$. Note that any test $T(Z^N)$ conducted by Willie satisfies~\cite[Theorem 13.1.1]{lehmann2006testing}
\begin{align}\label{eq:V-hypothesis}
  1\geq \alpha+\beta\geq 1-\V{\QcodeN}{\Qinn^{\otimes N}},
\end{align}
  where $\alpha$ and $\beta$ are the associated probabilities of false-alarm and missed-detection, respectively.

\begin{remark}
  Two important aspects of our model should be highlighted. First, we focus on random slot selection, in which codewords are placed in slots with known boundaries so that randomization happens at the codeword level. The studies in~\cite{Tchamkerten2009,Arumugam2016a} instead consider asynchronous communications, in which asynchronism happens at the symbol level. As shown in~\cite{Arumugam2016a}, the analysis of covertness in the latter case is possible but we were unable to obtain a precise characterization of the covert capacity in that case. Second, the study of sub-exponential random slot selection is of value because the randomization helps the transmitter hide its transmission; the randomization does not improve reliability but does not hurt it either because, as clarified in our proofs, the legitimate receiver can identify the slot in which the communication takes place.
\end{remark}

\begin{definition}
  \label{def:covertcapacity-subexp}
  Let $\delta\in(0,1)$. In the sub-exponential random slot selection regime $L_n=\omega(1)\cap \bigcap_{c>0} o(c^n)$, a covert throughput $r$ is achievable if there exists a sequence of codes with increasing blocklength as described above such that
  \begin{align}
    \lim_{n\to\infty}\frac{\log M}{\sqrt{n\log L_n}} \geq r,\quad \lim_{n\to\infty}P_e^{(n)}=0,\quad \lim_{n\to\infty}V^{(n)}\leq \delta.
  \end{align}
  The covert capacity $C_{\textnormal{covert}}$ is the supremum of achievable covert throughputs. 
\end{definition}

As often in covert communications, the choice of the normalization by $\sqrt{n\log L_n}$ shall be justified in hindsight when we develop a converse as part of the characterization of the covert capacity and obtain a finite value. 
For reference, we recall below the notion of covert capacity {without random slot selection}.

\begin{definition}
  \label{def:covertcapacity-sync}
  Let $\delta\in(0,1)$. Without random slot selection ($L_n=1$), a covert throughput $r$ is achievable if there exists a sequence of codes with increasing blocklength as described above such that
  \begin{align}
    \lim_{n\to\infty}\frac{\log M}{\sqrt{n}} \geq r,\quad \lim_{n\to\infty}P_e^{(n)}=0,\quad \lim_{n\to\infty}V^{(n)}\leq \delta.
  \end{align}
  The covert capacity $C_{\textnormal{covert}}$ is the supremum of achievable covert throughputs. 
\end{definition}

Our main results are then the following.

\begin{theorem}
  \label{thm:async-covert-capacity}
  In the sub-exponential random slot selection regime, if $\D{P_1}{P_0}>\D{Q_1}{Q_0}$, the covert capacity of a binary-input \ac{DMC} satisfies
  
  \begin{align}
    \frac{\D{P_1}{P_0}}{\sqrt{\chi_2(Q_1\Vert Q_0)}}\leq C_{\textnormal{covert}}  \leq \sqrt{\frac{2}{\chi_2(Q_1\Vert Q_0)}}\D{P_1}{P_0}.
  \end{align}
\end{theorem}
\begin{IEEEproof}
  See Section~\ref{sec:achv-DMC} and Section~\ref{sec:conv-DMC}.
\end{IEEEproof}
\begin{theorem}
  \label{thm:async-covert-capacity-Gaussian}
  In the sub-exponential random slot selection regime, if $\sigma_w^2>\sigma_b^2$, the covert capacity of an \ac{AWGN} channel satisfies

  \begin{align}
    \frac{\sigma_w^2}{\sqrt{2}\sigma_b^2}\leq C_{\textnormal{covert}}  \leq \frac{\sigma_w^2}{\sigma_b^2}.
  \end{align}
\end{theorem}
\begin{IEEEproof}
  The treatment of the \ac{AWGN} case requires separate proofs, see Section~\ref{sec:achv-AWGN} and Section~\ref{sec:conv-AWGN}.
\end{IEEEproof}
Recall that the covert capacity in the synchronous case is given by
\begin{align*}
  C_{\textnormal{covert}}=\frac{2Q^{-1}(\frac{1-\delta}{2})}{\sqrt{\chi_2(Q_1\Vert Q_0)}}\D{P_1}{P_0}
\end{align*}
for \acp{DMC}~\cite{Tahmasbi2019} and $C_{\textnormal{covert}}=\sqrt{2}Q^{-1}(\frac{1-\delta}{2})\frac{\sigma_w^2}{\sigma_b^2}$ for \ac{AWGN} channels~\cite{Zhang2019a,Wang2021}. Theorems~\ref{thm:async-covert-capacity} and~\ref{thm:async-covert-capacity-Gaussian} combined with the normalization \ac{wrt} $\sqrt{n\log L_n}$ introduced in Definition~\ref{def:covertcapacity-subexp} show that the benefits of introducing random slot selection is in boosting the scaling and circumventing the square root law. One subtlety is that random slot selection also replaces the dependence on the covertness constraint with a dependence on the number of slots, as seen when contrasting Definition~\ref{def:covertcapacity-subexp} and Definition~\ref{def:covertcapacity-sync} and noticing the change of normalization to $\sqrt{\log L_n}$. Also note that the dependence of the covert capacity on the parameter $\delta$ disappears with random slot selection.

The crux of our achievability proof lies in developing a bound through the relative entropy and Pinsker's inequality that captures the dependence of the covertness metric with the number of slots. Our converse proof follows standard practice and shows how the covertness constraint induces a weight constraint on the codewords, while again accurately capturing the dependence with the number of slots. Our results do not answer what happens in the unslotted asynchronism case, which as shown in~\cite{Arumugam2016a} is more difficult to analyze.

The relative simplicity of the results presented in Theorem~\ref{thm:async-covert-capacity} and Theorem~\ref{thm:async-covert-capacity-Gaussian} hides a more substantial difficulty in the proofs. The key challenge that we have addressed is how to precisely quantify the effect of the number of slots $L_n$ on the covert capacity. This effect is not straightforward to identify because it requires one to precisely analyze how randomization, creating a mixture of distributions $\widehat{Q}_{Z}^N$, helps reduce the distance between the mixture $\widehat{Q}_{Z}^N$ and the reference distribution $Q_0^{\otimes N}$. Our approach requires a fairly intricate back and forth between the analysis of relative entropy and total variation distance to make analytical progress towards the solution.

\section{Proofs of Theorem~\ref{thm:async-covert-capacity}}
\label{sec:proofs}

\subsection{Preliminary lemma}
\label{sec:prelim-DMC}

Our achievability proof shall rely on the following lemma adapted from our preliminary results in~\cite{Arumugam2016a} to bound the relative entropy.

\begin{lemma}
  \label{lm:prelim-dmc}
  Consider a binary-input \ac{DMC} with input $\set{x_0,x_1}$, transition probability $W_{Z|X}$, and output $\calZ$. Set $Q_0\eqdef W_{Z|X=x_0}$ and $Q_1\eqdef W_{Z|X=x_1}$. Let $\alpha>0$, $L\in\bbN_*$, and $N\eqdef nL$, and define the distribution $Q_{\alpha}$ such that $Q_{\alpha}(z)=(1-\alpha)Q_0(z)+\alpha Q_1(z)$ and $Q_{\alpha}^{N}\eqdef \E[T]{Q_{\alpha,T}^{N}}$ where
  \begin{align}\label{eq:DMC-Q-alpha-N}
    Q_{\alpha,t}^{N}(z^{N})\eqdef \left(\prod_{\ell=1}^{t-1}\Qinn^{\otimes n}(\bfz_{\intseq{n(\ell-1)+1}{n\ell}})\right)\left(Q_{\alpha}^{\otimes n}(\bfz_{\intseq{n(t-1)+1}{nt}})\right)\left(\prod_{\ell=t+1}^{L}\Qinn^{\otimes n}(\bfz_{\intseq{n(\ell-1)+1}{n\ell}})\right).
\end{align}
Then,
\begin{align}
  \D{Q^N_{\alpha}}{Q^{\otimes N}_0} \leq e^{n\alpha^2\chi_2(Q_1\Vert Q_0)-\log L}.
\end{align}
\end{lemma}

\begin{IEEEproof}
  We first define
  \begin{align*}
    \Psi(z)\eqdef\frac{Q_1(z)-Q_0(z)}{Q_0(z)}.
  \end{align*}
  Then,
  \begin{align}
&\D{Q^N_{\alpha}}{Q^{\otimes N}_0}\nonumber\\
&=\sum_{z^N}Q_0^{\otimes N}(z^N)\frac{1}{L}\sum_{\ell=1}^L\prod_{i=n(\ell-1)+1}^{n\ell}\frac{Q_{\alpha}(z_i)}{Q_0(z_i)}\log\left(\sum_{k=1}^L\frac{1}{L}\prod_{j=n(k-1)+1}^{nk}\frac{Q_{\alpha}(z_j)}{Q_0(z_j)}\right)\nonumber\\
&\labrel={eq:def-set-size}\mathbb{E}_{\Qinn^{\otimes N}}\left[\left(1+\frac{1}{L}\sum_{\ell=1}^L\sum_{m=1}^n\alpha^m\sum_{\calS_m\subseteq\intseq{n(\ell-1)+1}{n\ell}}\prod_{u=1}^m\Psi(Z_{\calS_{m,u}})\right)\right.\nonumber\\
    &\qquad\left.\times\log\left(1+\frac{1}{L}\sum_{k=1}^L\sum_{d=1}^n\alpha^d\sum_{\calS_d\subseteq\intseq{n(k-1)+1}{nk}}\prod_{w=1}^d\Psi(Z_{\calS_{d,w}})\right)\right]\nonumber\\
&\labrel\leq{eq:log-taylor}\frac{1}{L}\sum_{k=1}^L\sum_{d=1}^n\alpha^d\sum_{\calS_d\subseteq\intseq{n(k-1)+1}{nk}}\E[\Qinn^{\otimes N}]{\prod_{w=1}^d\Psi(z_{\calS_{d,w}})}\nonumber\\
&\quad+\frac{1}{L^2}\sum_{\ell,k=1}^L\sum_{m,d=1}^n\alpha^{m+d}\sum_{\substack{\calS_m\subseteq\intseq{n(\ell-1)+1}{n\ell}\\ \calS_d\subseteq\intseq{n(k-1)+1}{nk}}}\E[\Qinn^{\otimes N}]{\prod_{u=1}^m\Psi(z_{\calS_{m,u}})\prod_{w=1}^d\Psi(z_{\calS_{d,w}})}\nonumber\\
&\labrel={eq:DMC-achv-chi}\frac{1}{L}\sum_{m=1}^n\alpha^{2m}\chi_2(Q_1\Vert Q_0)^{m}\sum_{\calS_m\subseteq\intseq{n(\ell-1)+1}{n\ell}}1\nonumber\\
&=\frac{1}{L}\sum_{m=1}^n\binom{n}{m}\alpha^{2m}\chi_2(Q_1\Vert Q_0)^m\nonumber\\
&=\frac{(1+\alpha^2\chi_2(Q_1\Vert Q_0))^n-1}{L}\nonumber\\
    \label{eq:DMC-achv-D}
    &\labrel\leq{eq:upper-bound-binom}\frac{e^{n\alpha^2\chi_2(Q_1\Vert Q_0)}-1}{L}\leq\frac{e^{n\alpha^2\chi_2(Q_1\Vert Q_0)}}{L},
  \end{align}
  where \eqref{eq:def-set-size} follows by defining $\calS_m$ as an ordered subset of $\intseq{1}{N}$ with cardinality $m\in\bbN_*$, \eqref{eq:log-taylor} follows from $(1+x)\log(1+x)\leq x+x^2$ for $x>-1$, \eqref{eq:DMC-achv-chi} follows since $\E[Q_0]{\Psi(Z)}=0$, $\E[Q_0]{\Psi^2(Z)}=\chi_{2}(Q_1\Vert Q_0)$, and \eqref{eq:upper-bound-binom} follows from $1+x\leq e^x$ for $x>0$.
  
\end{IEEEproof}

\subsection{Achievability for \aclp{DMC}}
\label{sec:achv-DMC}
\begin{proposition}
  \label{prop:achv-DMC}
  Consider a binary-input \ac{DMC} with $\calX=\{x_0, x_1\}$ with $P_1\ll P_0$, $Q_1\ll Q_0$, and $Q_1\neq Q_0$. For any $\xi\in(0, 1)$, if $\D{P_1}{P_0}>\D{Q_1}{Q_0}$, then there exist covert communication schemes with sub-exponential random slot selection such that
  \begin{align}
    &\limn{\frac{\log M}{\sqrt{n\log L_n}}}=(1-\xi)\sqrt{\frac{1}{\chi_2(Q_1\Vert Q_0)}}\D{P_1}{P_0},\\
    &\limn{P_e^{(n)}}=0,~\limn{V^{(n)}}\leq\delta.
  \end{align}
\end{proposition}
\begin{IEEEproof}
Our proof is an extension of \cite{Arumugam2016a} that relies on random coding arguments, soft-covering techniques~\cite{Han1993,Cuff2013} (or, channel resolvability~\cite{Bloch2016}) and Bernstein-type concentration inequality.  
\paragraph{Codebook Construction}
We first generate $M\in\bbN_*$ random codewords $\{{\bfx}_w\}_{w=1}^M\in\{x_0, x_1\}^n$ independently according to the product distribution $\Pi_{\alpha_n}^{\otimes n}$, where $$\Pi_{\alpha_n}(x)\eqdef(1-\alpha_n)\indic{x=x_0}+\alpha_n\indic{x=x_1},$$ $\{\alpha_n\}_n>0$ is a sequence that will be determined later such that $\limn{\alpha_n}=0$. Note that $\Pi_{\alpha_n}$ therefore induces distributions $P_{\alpha_n}\eqdef(1-\alpha_n)P_0+\alpha_nP_1$ and $Q_{\alpha_n}\eqdef(1-\alpha_n)Q_0+\alpha_nQ_1$. Alice then chooses the message $w\in\intseq{1}{M}$ and a transmission slot $t\in\intseq{1}{L_n}$ uniformly at random to generate a sequence of channel input $x_{w,t}^N$.
By choosing
\begin{align*}
\alpha_n=\sqrt{\frac{\log \left(L_n\left(2\delta^2-4n^{-\frac{1}{2}}\right)\right)}{n\chi_2(Q_1\Vert Q_0)}},
\end{align*}
Lemma~\ref{lm:prelim-dmc} shows that
\begin{align*}
  \D{{Q}^{N}_{\alpha_n}}{\Qinn^{\otimes N}}\leq 2\delta^2-4n^{-\frac{1}{2}}.
\end{align*}
Consequently, with Pinsker's inequality we obtain
\begin{align}\label{eq:DMC-achv-covert-proc}
  \V{Q^N_{\alpha_n}}{\Qinn^{\otimes N}}\leq\delta-n^{-\frac{1}{2}}.
\end{align}
\paragraph{Channel Reliability Analysis}
We define the following decoding region:
  \begin{align}
    \calA_\gamma^n\eqdef\left\{(\bfx, \bfy)\in\calX^n\times\calY^n:\log\frac{\Wbn(\bfy|\bfx)}{P_{\alpha_n}^{\otimes n}(\bfy)}>\gamma\right\}.
  \end{align}
Bob's decoder operates as follows: the decoder checks one slot at a time and starts with slot $t=1$. If there exists a unique codeword such that $(\bfx_w, \bfy_{\intseq{n(t-1)+1}{nt}})\in\calA_\gamma^n$, then it outputs $\widehat{W}=w$ and stops decoding. Otherwise, it proceeds with the next time slot by incrementing $t$ and reiterates the same procedure. The decoding terminates at time slot $t=L_n$ if there is still no such $\widehat{W}$ and declares an error. 
\begin{lemma}
  \label{lm:DMC-ch-reliability}
  By choosing
  \begin{align}\label{eq:DMC-ch-rel-message-size}
    \log M=(1-\delta_1)(1-\nu_1)n(\alpha_n\D{\Pinf}{\Pinn}+\calO\left(\alpha_n^2\right)),
  \end{align}
  the expected average probability of error over the choice of random codebook satisfies
  \begin{align}\label{eq:DMC-ch-rel-err}
    \E[\calC]{P_e^{(n)}}\leq\exp\left(-\xi_1\sqrt{n\log L_n}\right),
  \end{align}
  for some $\nu_1,\delta_1\in(0, 1)$, $\xi_1>0$, and $n$ large enough.
\end{lemma}
\begin{IEEEproof}
  Provided in Appendix~\ref{sec:proof-DMC-ch-rel}.
\end{IEEEproof}
\paragraph{Soft-covering Analysis}
\begin{lemma}
  \label{lm:DMC-ch-rsl}
  By choosing
  \begin{align}\label{eq:DMC-ch-rsl-key-rate}
    \log M=(1+\delta_2)(1+\nu_2)n(\alpha_n\D{Q_1}{Q_0}+\calO(\alpha_n^2)),
  \end{align}
  the expected variational distance over the choice of random codebook between $\QcodeN$ and $Q_{\alpha_n}^{N}$ satisfies
  \begin{align}\label{eq:DMC-ch-rsl-V}
    \E[\calC]{\V{\QcodeN}{Q_{\alpha_n}^N}}\leq\exp\left(-\xi_2\sqrt{n\log L_n}\right)
  \end{align}
  for some $\nu_2,\delta_2\in(0,1)$, $\xi_2>0$, and $n$ large enough.
\end{lemma}
\begin{IEEEproof}
  Provided in Appendix~\ref{sec:proof-DMC-ch-rsl}.
\end{IEEEproof}
\paragraph{Codebook Identification}
Note that since $\D{P_1}{P_0}>\D{Q_1}{Q_0}$, we can always choose $\nu_1,\nu_2, \delta_1$ and $\delta_2$ arbitrarily small such that
\begin{align*}  (1-\delta_1)(1-\nu_1)n\left(\alpha_n\D{P_1}{P_0}+\calO(\alpha_n^2)\right)
  >(1+\delta_2)(1+\nu_2)n\left(\alpha_n\D{Q_1}{Q_0}+\calO(\alpha_n^2)\right)
\end{align*}
to satisfy \eqref{eq:DMC-ch-rel-message-size} and \eqref{eq:DMC-ch-rsl-key-rate} for $\log M$. Then, from \eqref{eq:DMC-ch-rel-err} and \eqref{eq:DMC-ch-rsl-V}, and by using Markov's inequality, we can identify at least one codebook over the random code generation with sufficiently large $n$ and some constants $\overline{\xi}_1>0$ and $\overline{\xi}_2>0$ such that
\begin{align}\label{eq:DMC-iden-err}
  P_e^{(n)}&\leq\exp\left(-\overline{\xi}_1\sqrt{n\log L_n}\right),\\
  \label{eq:DMC-iden-V}
  \V{\QcodeN}{Q_{\alpha_n}^{N}}&\leq\exp\left(-\overline{\xi}_2\sqrt{n\log L_n}\right).
\end{align}
Consequently, by combining~\eqref{eq:DMC-achv-covert-proc} and \eqref{eq:DMC-iden-V}, the triangle inequality implies that
\begin{align}
  \V{\QcodeN}{\Qinn^{\otimes N}}&\leq\V{\QcodeN}{Q_{\alpha_n}^{N}}+\V{Q_{\alpha_n}^N}{\Qinn^{\otimes N}}\nonumber\\
                                  &\leq\exp\left(-\overline{\xi}_2\sqrt{n\log L_n}\right)+\delta-n^{-\frac{1}{2}},\label{eq:triangleq}
\end{align}
and it guarantees the covertness constraint.
\paragraph{Throughput Analysis}
Now we can simply set $1-\xi=(1-\delta_1)(1-\nu_1)$ and obtain
\begin{align}
  &\limn{\frac{\log M}{\sqrt{n\log L_n}}}\nonumber\\&=\limn{\frac{(1-\xi)n\left(\alpha_n\D{P_1}{P_0}+\alpha_n^2\right)}{\sqrt{n \log L_n}}}\nonumber\\
                                        &=(1-\xi)\frac{\D{P_1}{P_0}}{\sqrt{\chi_2(Q_1\Vert Q_0)}}\limn{\sqrt{1+\frac{\log\left(\delta-4n^{-\frac{1}{2}}\right)}{\log L_n}}}\nonumber\\
  &=(1-\xi)\sqrt{\frac{1}{\chi_2(Q_1\Vert Q_0)}}\D{P_1}{P_0}.
\end{align}
\end{IEEEproof}
\subsection{Converse for \aclp{DMC}}
\label{sec:conv-DMC}
\begin{proposition}
  \label{prop:conv-DMC}
  Consider a binary-input \ac{DMC} with $\calX=\{x_0, x_1\}$, $P_1\ll P_0$, $Q_1\ll Q_0$, and $Q_1\neq Q_0$. If there exists a sequence of keyless covert communication schemes with sub-exponential random slot selections with increasing blocklength $n\in\bbN_*$, characterized by $\limn{P_e^{(n)}}=0,~\limn{V^{(n)}}\leq\delta$ and $\limn{M}=\infty$, then
  \begin{align}
    \limn{\frac{\log M}{\sqrt{n\log L_n}}}\leq\sqrt{\frac{2}{\chi_2(Q_1\Vert Q_0)}}\D{P_1}{P_0}.
  \end{align}
\end{proposition}
\begin{IEEEproof}
  Our converse proof is motivated by \cite{Goeckel2016,Bash2016,Dani2021Covert} and also borrows from \cite[Section IV.-B]{Tahmasbi2019}.
  \paragraph{Lower Bound on Covertness Metric}
  We first show with the following lemma that for any given code $\calM$, if the minimum weight of codewords satisfies $w_*>\sqrt{\frac{2n\log L_n}{\chi_2(Q_1\Vert Q_0)}}$, then it cannot be covert.
  \begin{lemma}
    \label{lm:DMC-conv-min-weight}
    Consider a specific codebook $\calM$ with increasing blocklength $n$. Let $w_*$ be the minimum weight of codewords within $\calM$, i.e., $w_*\eqdef\min_{\bfx\in\calM}\wt{\bfx}$. If $w_*>\sqrt{\frac{2n\log L_n}{\chi_2(Q_1\Vert Q_0)}},$ then both the probabilities of false alarm $\alpha$ and missed detection $\beta$ satisfy $\limn{\alpha}=0$ and $\limn{\beta}=0$.
  \end{lemma}
  \begin{IEEEproof}
    Provided in Section~\ref{sec:proof-DMC-conv-min-weight}.
  \end{IEEEproof}
  Then we conclude that, if $w_*>\sqrt{\frac{2n\log L_n}{\chi_2(Q_1\Vert Q_0)}}$, \eqref{eq:V-hypothesis} implies that
  \begin{align}\label{eq:DMC-conv-V-bound}
    \limn{V^{(n)}}&=\limn{\V{\widehat{Q}^N_Z}{Q^{\otimes N}_0}}\geq 1-\limn{\alpha}-\limn{\beta}=1.
  \end{align}
  \paragraph{Existence of a Good Sub-codebook}
We argue that in order for a codebook $\calC$ to be covert, there must exist a sub-codebook comprising a positive fraction of the original codebook that consists of low-weight codewords.
We consider the covert code of interest $\calC$ such that $\limn{V^{(n)}}\leq\delta$, and partition $\calC$ into a high-weight sub-codebook $\calC^{(h)}$ and a low-weight sub-codebook $\calC^{(\ell)}$. In particular, we define
\begin{align}
  \calC^{(\ell)}\eqdef\left\{\bfx\in\calC:\wt{\bfx}\leq\sqrt{\frac{2n\log L_n}{\chi_2(Q_1\Vert Q_0)}}\right\},
\end{align}
and $\calC^{(h)}\eqdef\calC\backslash \calC^{(\ell)}$. The two associated output distributions induced at Willie's terminal are $\widehat{Q}_Z^{N,(\ell)}\eqdef\E[T]{\widehat{Q}^{N,(\ell)}_{Z,T}}$ and $\widehat{Q}_Z^{N,(h)}\eqdef\E[T]{\widehat{Q}^{N,(h)}_{Z,T}}$, where
\begin{align*}
  \widehat{Q}_{Z,t}^{N,(\ell)}(z^N)&\eqdef\left(\prod_{k=1}^{t-1}Q_0^{\otimes n}(\bfz_{\intseq{n(k-1)+1}{nk}})\right)\left(\frac{1}{\card{\calC^{(\ell)}}}\sum_{\bfx\in\calC^{(\ell)}}\Wwn(\bfz|\bfx)\right)\left(\prod_{k=t+1}^{L_n}Q_0^{\otimes n}(\bfz_{\intseq{n(k-1)+1}{nk}})\right),~\text{and}\\
\widehat{Q}_{Z,t}^{N,(h)}(z^N)&\eqdef\left(\prod_{k=1}^{t-1}Q_0^{\otimes n}(\bfz_{\intseq{n(k-1)+1}{nk}})\right)\left(\frac{1}{\card{\calC^{(h)}}}\sum_{\bfx\in\calC^{(h)}}\Wwn(\bfz|\bfx)\right)\left(\prod_{k=t+1}^{L_n}Q_0^{\otimes n}(\bfz_{\intseq{n(k-1)+1}{nk}})\right),
\end{align*}
respectively. Note that $\QcodeN=\frac{\card{\calC^{(h)}}}{\card{\calC}}\widehat{Q}^{N,(h)}_{Z}+\frac{\card{\calC^{(\ell)}}}{\card{\calC}}\widehat{Q}^{N,(\ell)}_{Z}$. Then,
\begin{align}
  V^{(n)}&=\V{\QcodeN}{Q_0^{\otimes N}}\nonumber\\
         &\labrel\geq{eq:DMC-conv-V-tri}\frac{\card{\calC^{(h)}}}{\card{\calC}}\V{\widehat{Q}^{N,(h)}_Z}{Q_0^{\otimes N}}-\frac{\card{\calC^{(\ell)}}}{\card{\calC}}\V{\widehat{Q}^{N,(\ell)}_Z}{Q_0^{\otimes N}}\nonumber\\
         &=\V{\widehat{Q}^{N,(h)}_Z}{Q_0^{\otimes N}}-\frac{\card{\calC^{(\ell)}}}{\card{\calC}}\left[\V{\widehat{Q}^{N,(h)}_Z}{Q_0^{\otimes N}}+\V{\widehat{Q}^{N,(\ell)}_Z}{Q_0^{\otimes N}}\right]\nonumber\\
         &\labrel\geq{eq:DMC-conv-V-var-lower-1}\V{\widehat{Q}^{N,(h)}_Z}{Q_0^{\otimes N}}-2\frac{\card{\calC^{(\ell)}}}{\card{\calC}},
\end{align}
where \eqref{eq:DMC-conv-V-tri} follows since
\begin{align*}
  &\frac{\card{\calC^{(h)}}}{\card{\calC}}\V{\widehat{Q}^{N,(h)}_Z}{Q_0^{\otimes N}}\\
  &=\frac{1}{2}\norm[1]{\left(\QcodeN-\Qinn^{\otimes N}\right)-\frac{\card{\calC^{(\ell)}}}{\card{\calC}}\left(\widehat{Q}^{N,(\ell)}_Z-\Qinn^{\otimes N}\right)}\\
  &\leq\V{\QcodeN}{\Qinn^{\otimes N}}+\frac{\card{\calC^{(\ell)}}}{\card{\calC}}\V{\widehat{Q}^{N,(\ell)}_Z}{\Qinn^{\otimes N}},
\end{align*}
and \eqref{eq:DMC-conv-V-var-lower-1} follows from the fact that variational distance is less than $1$.
Since \eqref{eq:DMC-conv-V-bound} implies that
\begin{align*}
  \limn{\V{\widehat{Q}^{N,(h)}_Z}{\Qinn^{\otimes N}}}\geq 1,
\end{align*}
we obtain that
\begin{align*}
  \delta&\geq\limn{V^{(n)}}\geq\limn{\V{\widehat{Q}^{N,(h)}_Z}{\Qinn^{\otimes N}}}-2\limn{\frac{\card{\calC^{\ell}}}{\card{\calC}}}\geq 1-2\limn{\frac{\card{\calC^{\ell}}}{\card{\calC}}},
\end{align*}
 and it consequently implies that
 \begin{align}
   \limn{\frac{\card{\calC^{(\ell)}}}{\card{\calC}}}\geq\frac{1-\delta}{2},
 \end{align}
 which asserts that a non-vanishing fraction of the codewords within $\calC$ must have weights less than $\sqrt{\frac{2n\log L_n}{\chi_2(Q_1\Vert Q_0)}}$ for $n$ large enough.
\paragraph{Upper Bound on Covert Throughput}
Let $M=\card{\calC}$. For any given $\kappa\in(0, \frac{1-\delta}{2})$, there exists some $N>0$ depending on $\kappa$ such that for $n>N$, $\card{\calC^{(\ell)}}\geq(\frac{1-\delta}{2}-\kappa)M$. Furthermore, since the average probability of error of $\calC$ is at most $P_e^{(n)}$, the average probability of error of $\calC^{\ell}$ is at most $\frac{2}{1-\delta-2\kappa}P_e^{(n)}$. Let $\overline{W}$ denote the uniformly distributed message over $\calC^{\ell}$ and assume a slightly stronger decoder that has the knowledge of the transmission time. Following standard techniques, we have
\begin{align}\label{eq:DMC-conv-V-sub-code-size}
  \log\card{\calC^{(\ell)}}&=\avgH{\overline{W}}\nonumber\\
                           &\leq\avgI{\overline{W};\bfY}+\avgH{\overline{W}|\bfY}\nonumber\\
                           &\leq\avgI{\bfX;\bfY}+\frac{2P_e^{(n)}}{1-\delta-2\kappa}\log\card{\calC^{(\ell)}}+\Hb{\frac{2P_e^{(n)}}{1-2\kappa}}\nonumber\\
  &\leq n\avgI{\overline{X};\overline{Y}}+\frac{2P_e^{(n)}}{1-\delta-2\kappa}\log\card{\calC^{(\ell)}}+1,
\end{align}
where the random variables $\overline{X}$ and $\overline{Y}$ follow
\begin{align*}
  \Pi_{\overline{X}}(x)\eqdef\frac{1}{n}\sum_{i=1}^n\Pi_{X_i}(x)=\frac{1}{n\card{\calC^{(\ell)}}}\sum_{i=1}^n\sum_{\bfx\in\calC^{(\ell)}}\indic{x=x_i},
\end{align*}
and $\Wb\circ\Pi_{\overline{X}}$, respectively. Then, by \cite[Lemma 1]{Bloch2016},
\begin{align}
  \label{eq:DMC-conv-V-I} n\avgI{\overline{X};\overline{Y}}&\leq\max_{\bfx\in\calC^{(\ell)}}\wt{\bfx}\D{P_1}{P_0}\leq\sqrt{\frac{2n\log L_n}{\chi_2(Q_1\Vert Q_0)}}\D{P_1}{P_0}.
\end{align}
By combining \eqref{eq:DMC-conv-V-sub-code-size} with \eqref{eq:DMC-conv-V-I}, we have
\begin{align}
  \log\card{\calC^{(\ell)}}\leq\frac{\sqrt{\frac{2n\log L_n}{\chi_2(Q_1\Vert Q_0)}}\D{P_1}{P_0}+1}{1-\frac{2P_e^{(n)}}{1-\delta-2\kappa}},
\end{align}
 and consequently, by taking $\kappa$ to be arbitrarily small and $n$ large enough,
 \begin{align}
   \limn{\frac{\log M}{\sqrt{n\log L_n}}}&\leq\limn{\frac{\log\card{\calC^{(\ell)}}-\log\left(\frac{1-\delta}{2}-\kappa\right)}{\sqrt{n\log L_n}}}\nonumber\\
   &=\sqrt{\frac{2}{\chi_2(Q_1\Vert Q_0)}}\D{P_1}{P_0}.
 \end{align}
\end{IEEEproof}

\section{Proof of Theorem~\ref{thm:async-covert-capacity-Gaussian}}
\label{sec:proof-thm-Gaussian}

\subsection{Preliminary Lemma}
\label{sec:prelim-AWGN}
Our achievability proof for \ac{AWGN} channels shall rely on the following lemma to bound the relative entropy.
\begin{lemma}
  \label{lm:prelim-AWGN}
  Consider the \ac{AWGN} channel model in~\eqref{eq:AWGN-model}. Let $a>0$, $L\in\bbN_*$, $\rho\eqdef a^2$, and $N\eqdef nL$. Define the distributions $Q_0,Q_{\rho}$ with densities $q_0,q_\rho$ such that $Q_0=\calN(0, \sigma_w^2)$ and $Q_{\rho}(z)=\frac{1}{2}\calN(a, \sigma_w^2)+\frac{1}{2}\calN(-a, \sigma_w^2)$, respectively. Define $Q_{\rho}^{N}\eqdef \E[T]{Q_{\alpha,T}^{N}}$ where
  \begin{align}\label{eq:AWGN-Q-rho-n}
    Q_{\rho,t}^{N}(z^{N})\eqdef \left(\prod_{\ell=1}^{t-1}\Qinn^{\otimes n}(\bfz_{\intseq{n(\ell-1)+1}{n\ell}})\right)\left(Q_{\rho}^{\otimes n}(\bfz_{\intseq{n(t-1)+1}{nt}})\right)\left(\prod_{\ell=t+1}^{L}\Qinn^{\otimes n}(\bfz_{\intseq{n(\ell-1)+1}{n\ell}})\right).
\end{align}
Then,
\begin{align}
  \D{q^N_{\rho}}{q^{\otimes N}_0} \leq e^{\frac{n\rho^2}{2\sigma_w^4}-\log L}.
\end{align}
\end{lemma}
\begin{IEEEproof}
  
  Note that
\begin{align*}
  &q^{\otimes N}_{\rho}(z^N)\nonumber\\&
  \eqdef\frac{1}{L}\sum_{\ell=1}^L\prod_{i\in\intseq{1}{n(\ell-1)}}q_0(z_i)\prod_{i\in\intseq{n(\ell-1)+1}{n\ell}}q_\rho(z_i)\prod_{i\in\intseq{n\ell+1}{nL}}q_0(z_i)\\
&=q_0^{\otimes N}(z^N)\frac{1}{L}\sum_{\ell=1}^L\prod_{i\in\intseq{n(\ell-1)+1}{n\ell}}\frac{q_\rho(z_i)}{q_0(z_i)}\\
&=q_0^{\otimes N}(z^N)\frac{1}{L}\sum_{\ell=1}^L\prod_{i\in\intseq{n(\ell-1)+1}{n\ell}}\left(1+\frac{(q_\rho(z_i)-q_0(z_i))}{q_0(z_i)}\right)\\
&=q_0^{\otimes N}(z^N)\left(1+\frac{1}{L}\sum_{\ell=1}^L\sum_{m=1}^n\sum_{\calS_m\subseteq\intseq{n(\ell-1)+1}{n\ell}}\prod_{u=1}^mA(z_{\calS_{m,u}})\right),
\end{align*}
where $A(z)\eqdef\frac{q_\rho(z)-q_0(z)}{q_0(z)}$ and $\calS_m$ is an ordered subset of $\intseq{1}{N}$ with cardinality $m\in\bbN_*$.
Similar to \eqref{eq:DMC-achv-D},
\begin{align}\label{eq:AWGN-achv-D-1}
&\D{q_\rho^{\otimes N}}{q_0^{\otimes N}}\nonumber\\
&=\mathbb{E}_{q_0^{\otimes N}}\left[\frac{1}{L}\sum_{\ell=1}^L\prod_{i=n(\ell-1)+1}^{n\ell}\frac{q_\rho(z_i)}{q_0(z_i)}\log\left(\sum_{k=1}^L\frac{1}{L}\prod_{j=n(k-1)+1}^{nk}\frac{q_\rho(z_j)}{q_0(z_j)}\right)\right]\nonumber\\
&=\mathbb{E}_{q_0^{\otimes N}}\left[\left(1+\frac{1}{L}\sum_{\ell=1}^L\sum_{m=1}^n\sum_{\calS_m\subseteq\intseq{n(\ell-1)+1}{n\ell}}\prod_{u=1}^mA(z_{\calS_{m, u}})\right)\right.\nonumber\\
  &\left.\phantom{====}\times\log\left(1+\frac{1}{L}\sum_{k=1}^L\sum_{d=1}^n\sum_{\calS_d\subseteq\intseq{n(k-1)+1}{nk}}\prod_{w=1}^dA(z_w)\right)\right]\nonumber\\
&\labrel\leq{eq:AWGN-D-log-bound}\frac{1}{L}\sum_{k=1}^L\sum_{d=1}^n\sum_{\calS_d\subseteq\intseq{n(k-1)+1}{nk}}\E[q_0^{\otimes N}]{\prod_{w=1}^dA_n(z_{\calS_{d,w}})}\nonumber\\
&\quad+\frac{1}{L^2}\sum_{\ell=1}^L\sum_{k=1}^L\sum_{m=1}^n\sum_{d=1}^n\sum_{\substack{\calS_m\subseteq\intseq{n(\ell-1)+1}{n\ell}\\ \calS_d\subseteq\intseq{n(k-1)+1}{nk}}}\E[q_0^{\otimes N}]{\prod_{u=1}^mA_n(z_{\calS_{m,u}})\prod_{w=1}^dA_n(z_{\calS_{d,w}})}\nonumber\\
&\labrel={eq:AWGN-D-chi}\frac{1}{L}\sum_{m=1}^n\left(\cosh(\frac{\rho}{\sigma_w^2})-1\right)^m\sum_{\calS_m\subseteq\intseq{n(\ell-1)+1}{n\ell}}1\nonumber\\
&=\frac{\cosh(\frac{\rho}{\sigma_w^2})^n-1}{L}\nonumber\\
  &\labrel\leq{eq:AWGN-D-cosh}\frac{e^{\frac{n\rho^2}{2\sigma_w^4}}-1}{L}\leq\frac{e^{\frac{n\rho^2}{2\sigma_w^4}}}{L},
\end{align}
where \eqref{eq:AWGN-D-log-bound} follows from $(1+x)\log(1+x)\leq x+x^2$ for $x>-1$, \eqref{eq:AWGN-D-chi} follows since
\begin{align*}
  \chi_1^{(n)}&\eqdef\E[Q_0]{A(Z)}=0,\\
  \chi_2^{(n)}&\eqdef\E[Q_0]{A^2(Z)}=\frac{1}{2}\left(e^{\frac{a^2}{\sigma_w^2}}+e^{-\frac{a^2}{\sigma_w^2}}\right)-1=\cosh(\frac{\rho}{\sigma_w^2})-1,
\end{align*}
and \eqref{eq:AWGN-D-cosh} follows from $\cosh(x)\leq\exp(\frac{x^2}{2})$ for $x\in\bbR$.
\end{IEEEproof}
\subsection{Achievability for \acl{AWGN} Channels}
\label{sec:achv-AWGN}
\begin{proposition}
  \label{prop:achv-AWGN}
  Consider the \ac{AWGN} channel model in~\eqref{eq:AWGN-model}. For any $\xi\in(0, 1)$, if $\sigma_w^2>\sigma_b^2$, then there exist covert communication schemes with sub-exponential random slot selection such that
  \begin{align}
    &\limn{\frac{\log M}{\sqrt{n\log L_n}}}=(1-\xi)\frac{\sigma_w^2}{\sqrt{2}\sigma_b^2},\\
    &\limn{P_e^{(n)}}=0, \limn{V^{(n)}}\leq\delta.
  \end{align}
\end{proposition}
\begin{IEEEproof}
  The proof largely follows the ones given in \cite{Zhang2019a,Wang2021} to construct a random codebook based on \ac{BPSK} with constellation points vanishing to the origin with blocklength, and we reiterate key steps similar to those introduced in Section~\ref{sec:achv-DMC}.
\paragraph{Codebook Construction}
We let $x_0=0$ be the innocent symbol. Alice generates $M\in\bbN_*$ random \ac{BPSK} codewords $\{\bfx_w\}_{w=1}^M\in\{-a_n, a_n\}^{n}$ independently according to the product distribution $\Pi_{\rho_n}^{\otimes n}$, where
\begin{align*}
  \Pi_{\rho_n}(x)\eqdef\frac{1}{2}\indic{x=-a_n}+\frac{1}{2}\indic{x=a_n},
\end{align*}
$\{\rho_n\}$ is a sequence that will be determined later such that $\limn{\rho_n}=0$ and $\rho_n=\abs{a_n}^2$. Note that $\Pi_{\rho_n}$ therefore induces distributions $P_{\rho_n}\eqdef\Wb\circ\Pi_{\rho_n}=\frac{1}{2}\calN(-a_n, \sigma_b^2)+\frac{1}{2}\calN(a_n, \sigma_b^2)$ and $Q_{\rho_n}\eqdef\Ww\circ\Pi_{\rho_n}=\frac{1}{2}\calN(-a_n, \sigma_w^2)+\frac{1}{2}\calN(a_n, \sigma_w^2)$. Alice then chooses the message $w\in\intseq{1}{M}$ and a transmission slot $t\in\intseq{1}{L_n}$ uniformly at random to generate a sequence of channel input $x^N_{w,t}$. By choosing
\begin{align}\label{eq:AWGN-achv-rho}
  \rho_n=\sqrt{\frac{2\sigma_w^4\log \left(L_n\left(2\delta^2-4n^{-\frac{1}{2}}\right)\right)}{n}},
\end{align}
 Lemma~\ref{lm:prelim-AWGN} shows that
\begin{align*}
  \D{Q^N_{\rho_n}}{\Qinn^{\otimes N}}\leq 2\delta^2-4n^{-\frac{1}{2}}.
\end{align*}
Consequently, with Pinsker's inequality we obtain
\begin{align}\label{eq:AWGN-achv-cvt-proc}
  \V{Q^N_{\rho_n}}{\Qinn^{\otimes N}}\leq\delta-n^{-\frac{1}{2}}.
\end{align}
\paragraph{Channel Reliability Analysis}
We define the following decoding region:
  \begin{align*}
    \overline{\calA}_\gamma^n\eqdef\left\{(\bfx, \bfy)\in\calX^n\times\calY^n:\log\frac{\Wbn(\bfy|\bfx)}{\Pinn^{\otimes n}(\bfy)}>\gamma\right\},
  \end{align*}
and we use the same decoding principle as the one in Section~\ref{sec:achv-DMC}. Note that we use $\Pinn$ to define the decoding region instead of $P_{\rho_n}$. The main reason is to simplify the analysis of information density while incurring a small penalty as explained in Appendix~\ref{sec:proof-AWGN-ch-rel}.
\begin{lemma}
  \label{lm:AWGN-ch-rel}
  By choosing
  \begin{align}
    \label{eq:AWGN-ch-rel-message-size}
    \log M = (1-\delta_1)(1-\nu_1)n\left(\frac{\rho_n}{2\sigma_b^2}\right),
  \end{align}
  the expected average probability of error over the choice of random codebook satisfies
  \begin{align}\label{eq:AWGN-ch-rel-err}
    \E[\calC]{P_e^{(n)}}\leq\exp\left(-\theta_1\sqrt{n\log L_n}\right),
  \end{align}
  for some $\nu_1,\delta_1\in(0, 1)$ and $n$ large enough.
\end{lemma}
\begin{IEEEproof}
  Provided in Appendix~\ref{sec:proof-AWGN-ch-rel}.
\end{IEEEproof}
\paragraph{Soft-covering Analysis }
\begin{lemma}\label{lm:AWGN-ch-rsl}
  By choosing
  \begin{align}
    \label{eq:AWGN-ch-rsl-key-rate}
    \log M=(1+\delta_2)(1+\nu_2)n\frac{\rho_n}{2\sigma_w^2},
  \end{align}
  the expected variational distance over the choice of random codebook between $\QcodeN$ and $Q_{\rho_n}^N$ satisfies
  \begin{align}
    \label{eq:AWGN-ch-rsl-V}
    \E[\calC]{\V{\QcodeN}{Q_{\rho_n}^N}}\leq\exp\left(-\theta_2\sqrt{n\log L_n}\right)
  \end{align}
  for some $\nu_2,\delta_2\in(0, 1)$, $\theta_2>0$, and $n$ large enough.
\end{lemma}
\begin{IEEEproof}
  Provided in Appendix~\ref{sec:proof-AWGN-ch-rsl}.
\end{IEEEproof}
\paragraph{Identification of Codebook}
Note that since $\sigma_w^2>\sigma_b^2$, we can always choose $\nu_1,\nu_2,\delta_1$ and $\delta_2$ arbitrarily small such that
\begin{align*}
  (1-\delta_1)(1-\nu_1)n\frac{\rho_n}{2\sigma_b^2}>(1+\delta_2)(1+\nu_2)n\frac{\rho_n}{2\sigma_w^2}
\end{align*}
to satisfy \eqref{eq:AWGN-ch-rel-message-size} and \eqref{eq:AWGN-ch-rsl-key-rate} for $\log M$. Then, from \eqref{eq:AWGN-ch-rel-err} and \eqref{eq:AWGN-ch-rsl-V}, and by using Markov's inequality, we can identify at least one codebook over the random codebook generation with $n$ large enough and some constants $\bar{\theta}_1>0$ and $\bar{\theta}_2>0$ such that
\begin{align}
  P_e^{(n)}&\leq\exp\left(-\bar{\theta}_1\sqrt{n\log L_n}\right),\\
  \label{eq:AWGN-Q-iden-V}
  \V{\QcodeN}{Q_{\rho_n}^N}&\leq\exp\left(-\bar{\theta}_2\sqrt{n\log L_n}\right).
\end{align}
Then, by combining \eqref{eq:AWGN-achv-cvt-proc} with~\eqref{eq:AWGN-Q-iden-V}, the triangle inequality implies that
\begin{align}
  \V{\QcodeN}{\Qinn^{\otimes N}}&\leq\V{\QcodeN}{Q_{\rho_n}^{N}}+\V{Q_{\rho_n}^N}{\Qinn^{\otimes N}}\nonumber\\
  &\leq\exp\left(-\bar{\theta}_2\sqrt{n\log L_n}\right)+\delta-n^{-\frac{1}{2}},
\end{align}
which vanishes with increasing $n$ and, therefore, guarantees the covertness constraint. 
\paragraph{Throughput Analysis}
Choose $\xi>0$ such that $1-\xi=(1-\delta_1)(1-\nu_1)$ to obtain
\begin{align}
  &\quad\limn{\frac{\log M}{\sqrt{n\log L_n}}}=\limn{\frac{(1-\xi)n\frac{\rho_n}{2\sigma_b^2}}{\sqrt{n\log L_n}}}\nonumber\\
                                   &=(1-\xi)\frac{\sigma_w^2}{\sqrt{2}\sigma_b^2}\limn{\sqrt{1+\frac{\log\left(2\delta^2-4n^{-\frac{1}{2}}\right)}{\log L_n}}}\nonumber\\
  &=(1-\xi)\frac{\sigma_w^2}{\sqrt{2}\sigma_b^2}.
\end{align}
\end{IEEEproof}
\subsection{Converse for \acl{AWGN} Channels}
\label{sec:conv-AWGN}
\begin{proposition}
  \label{prop:conv-AWGN}
  Consider the \ac{AWGN} channel model in~\eqref{eq:AWGN-model}. If there exists a sequence of keyless covert communication schemes with sub-exponential random slot selection with increasing blocklength $n\in\bbN_*$, characterized by $\limn{P_e^{(n)}}=0,\limn{V^{(n)}}\leq\delta$ and $\limn{M}=\infty$, then
  \begin{align}
    \limn{\frac{\log M}{\sqrt{n\log L_n}}}\leq\frac{\sigma_w^2}{\sigma_b^2}.
  \end{align}
\end{proposition}
\begin{IEEEproof}
  Our converse proof borrows from \cite{Zhang2019a,Wang2021} and Section~\ref{sec:conv-DMC} and can be regarded as a refinement of \cite{Goeckel2016}.
\paragraph{Lower bound on Covertness Metric}
We first show in the following lemma that for any given codebook $\calM$, if the minimum power of codewords satisfies $P_*>\sqrt{4\sigma_w^4n\log L_n}$, then it cannot be covert.
\begin{lemma}\label{lm:AWGN-conv-V-min-power}
  Consider a specific codebook $\calM$ with increasing blocklength $n$. Let $P_*$ be the minimum power of codewords within $\calM$, i.e., $P_*\eqdef\min_{\bfx\in\calM}\norm[2]{\bfx}^2$. If
  \begin{align*}
    P_*>\sqrt{4\sigma_w^4n\log L_n},
  \end{align*}
  then both the probabilities of false-alarm $\alpha$ and missed-detection $\beta$ satisfy $\limn{\alpha}=0$ and $\limn{\beta}=0$.
\end{lemma}
\begin{IEEEproof}
  Provided in Section~\ref{sec:proof-AWGN-conv-V-min-power}.
\end{IEEEproof}
Then we conclude that, if $P_*>\sqrt{4\sigma_w^4n\log L_n}$, \eqref{eq:V-hypothesis} implies that
\begin{align}
  \limn{V^{(n)}}=\limn{\V{\QcodeN}{Q_0^{\otimes N}}}\geq 1-\limn{\alpha}-\limn{\beta}=1.
\end{align}
\paragraph{Existence of a Good Sub-codebook}
 Similar to the argument in Section~\ref{sec:conv-DMC}, for a codebook $\calC$ to be covert, i.e., $\limn{V^{(n)}}\leq\delta$, there is a non-vanishing fraction of the codewords within $\calC$ with power less than $\sqrt{4\sigma_w^4n\log L_n}$ for $n$ large enough. More formally, if we define
 \begin{align}\label{eq:AWGN-conv-V-def-low-power}
   \calC^{(\ell)}\eqdef\left\{\bfx\in\calC:\norm[2]{\bfx}^2\leq\sqrt{4\sigma_w^4n\log L_n}\right\},
 \end{align}
 then one can show that
 \begin{align}
   \limn{\frac{\card{\calC^{(\ell)}}}{\card{\calC}}}\geq\frac{1-\delta}{2}.
 \end{align}
\paragraph{Upper Bound on Covert Throughput}
Let $M=\card{\calC}$. For any given $\kappa\in(0,\frac{1-\delta}{2})$, there exists some $N>0$ depending on $\kappa$ such that for $n>N$, $\card{\calC^{(\ell)}}\geq(\frac{1-\delta}{2}-\kappa)M$. Furthermore, since the average probability of error of $\calC$ is at most $P_e^{(n)}$, the average probability of error of $\calC^{(\ell)}$ is at most $\frac{2}{1-\delta-2\kappa}P_e^{(n)}$. Let $\overline{W}$ denote the uniformly distributed message over $\calC^{(\ell)}$ and assume a slightly stronger decoder that has the knowledge of the transmission time. By the same standard techniques as the ones in~\eqref{eq:DMC-conv-V-sub-code-size},
\begin{align}
  \label{eq:AWGN-conv-V-sub-code-size}
  \log\card{\calC^{(\ell)}}\leq \frac{n\avgI{\overline{X};\overline{Y}}+1}{1-\frac{2P_e^{(n)}}{1-\delta-2\kappa}},
\end{align}
where the random variables $\overline{X}$ and $\overline{Y}$ follow
\begin{align*}
  \Pi_{\overline{X}}(x)\eqdef\frac{1}{n}\sum_{ui=1}^n\Pi_{X_i}(x)=\frac{1}{n\card{\calC^{(\ell)}}}\sum_{i=1}^n\sum_{\bfx\in\calC^{(\ell)}}\indic{x=x_i},
\end{align*}
 and $\Wb\circ \Pi_{\overline{X}}$. Let $\E{\overline{X}^2}=\overline{R}$, and consequently, $\E{\overline{Y}^2}=\overline{R}+\sigma_b^2$. Then,
 \begin{align}\label{eq:AWGN-conv-V-mutual-info}
   \avgI{\overline{X};\overline{Y}}&=\h{\overline{Y}}-\h{\overline{Y}|\overline{X}}\nonumber\\
                                   &\leq\frac{1}{2}\log\left(1+\frac{\overline{R}}{\sigma_b^2}\right)\nonumber\\
   &\labrel\leq{eq:AWGN-conv-V-log}\frac{\overline{R}}{2\sigma_b^2}\labrel\leq{eq:AWGN-conv-V-low-power}\frac{\sigma_w^2}{\sigma_b^2}\sqrt{\frac{\log L_n}{n}},
 \end{align}
 where \eqref{eq:AWGN-conv-V-log} follows from $\log(1+x)\leq x$ for all $x>0$ and \eqref{eq:AWGN-conv-V-low-power} follows from the definition of $\calC^{(\ell)}$. By combining \eqref{eq:AWGN-conv-V-sub-code-size} with \eqref{eq:AWGN-conv-V-mutual-info}, we obtain
 \begin{align}
   \log\card{\calC^{(\ell)}}\leq \frac{\frac{\sigma_w^2\sqrt{n\log L_n}}{\sigma_b^2}+1}{1-\frac{2P_e^{(n)}}{1-\delta-2\kappa}},
 \end{align}
 and consequently, by taking $\kappa$ to be arbitrarily small and $n$ large enough,
 \begin{align}
   \limn{\frac{\log M}{\sqrt{n\log L_n}}}\leq\limn{\frac{\log\card{\calC^{(\ell)}}-\log\left(\frac{1-\delta}{2}-\kappa\right)}{\sqrt{n\log L_n}}}=\frac{\sigma_w^2}{\sigma_b^2}.
 \end{align}
\end{IEEEproof}

\section{Concluding Remarks and Extensions}
\label{sec:concluding-remark}

{We conclude by discussing remaining challenges and relatively direct extensions of our results.}

\subsection{Closing the Gap between our Capacity Bounds}
\label{sec:rmk-bound-tight}

Our converse argument relies on the union bound to properly bound the probability of false-alarm and the maximum of test results over all slots. This is closely related to maximal inequalities~\cite[Section 2.5]{boucheronConcentrationInequalitiesNonasymptotic2013}, and we are not aware of any other techniques suitable to refine our approach. We conjecture that our converse is tight, as further supported by our analysis in Section~\ref{sec:rmk-secret-key}.

In contrast, our achievability relies on Lemma~\ref{lm:prelim-dmc} and Lemma~\ref{lm:prelim-AWGN} to bound the relative entropy and we conjecture that the bound could be refined to recover the desired $\sqrt{2}$ factor. Indeed, since Lemma~\ref{lm:prelim-dmc} does not require that $\alpha$ be small or rely on Taylor's expansion, our result would also apply to \emph{exponential} random slot selection, in which case our result would read $\log L_n>n\alpha^2\chi_2(Q_1\Vert Q_0)$. Compared to the requirement $\log L_n>n\D{Q_\alpha}{Q_0}$ obtained in~\cite[Theorem 1]{Dani2021Covert}, our result is looser since $\alpha^2\chi_2(Q_1\Vert Q_0)\geq\D{Q_{\alpha}}{Q_0}$ for any $\alpha\geq 0$. 
Unfortunately, we did not succeed in extending the approach of~\cite{Dani2021Covert} as it relies on fine control of the composition of codewords to invoke the method of type and extend techniques from \cite{Chandar2013}. In our sub-exponential random slot selection regime, the types are governed by $\alpha_n$ and the composition of codewords depends on $n$. It is also unclear to us how the approach can be extended to \ac{AWGN} channels.

\subsection{Covertness with Relative Entropy Metric and Secret Keys Required}
\label{sec:rmk-secret-key}

In light of the characterization of covert capacity developed in~\cite{Bloch2016}, it is legitimate to wonder if the condition $\D{Q_1}{Q_0}<\D{P_1}{P_0}$ for \acp{DMC} (or $\sigma_b^2<\sigma_w^2$ for \ac{AWGN} channels) is necessary for our theorems to hold, and how many secret keys bits are required if the condition is not satisfied. We provide an answer that parallels Theorem~\ref{thm:async-covert-capacity} but assumes that covertness is measured in terms of relative entropy. Specifically, we now assume that Alice and Bob both have access to a secret key $S\in\intseq{1}{K}$ unknown to Willie, which is used by Alice to encode and Bob to decode the message, respectively. Setting $P_{\textnormal{max}}^{(n)}=\max_{s}\E{\P{\smash{W\neq\widehat{W}}}|S=s}$ and $D^{(n)}\eqdef \D{\smash{\QcodeN}}{\Qinn^{\otimes N}}$, we modify the notion of achievable throughput as follows.
  \begin{definition}
    \label{def:modified_achievable}
      Let $\delta\in(0,1)$. In the sub-exponential random slot selection regime $L_n=\omega(1)\cap \bigcap_{c>0} o(c^n)$, a covert throughput $r$ is achievable under a relative entropy metric with associated key throughput $r_k$ if there exists a sequence of codes with increasing blocklength as described above such that
  \begin{align}
    &\lim_{n\to\infty}\frac{\log M}{\sqrt{n\log L_n}} \geq r,\quad \lim_{n\to\infty}P_{\textnormal{max}}^{(n)}=0,\\
    &\lim_{n\to\infty}\frac{\log K}{\sqrt{n\log L_n}} \leq r_k\quad \lim_{n\to\infty}D^{(n)}\leq \delta.
  \end{align}
  The covert capacity $C^{\textnormal{KL}}_{\textnormal{covert}}$ is the supremum of achievable covert throughputs, where the superscript $\textnormal{KL}$ indicates the change of covertness metric.
\end{definition}
We then have the following counterpart to Theorem~\ref{thm:async-covert-capacity}.
  \begin{theorem}
    \label{th:refinement}
    In the sub-exponential random slot selection regime, the covert capacity under relative entropy metric of a binary-input \ac{DMC} satisfies
  \begin{align}
    \frac{\D{P_1}{P_0}}{\sqrt{\chi_2(Q_1\Vert Q_0)}}\leq C^{\textnormal{KL}}_{\textnormal{covert}}  \leq \sqrt{\frac{2}{\chi_2(Q_1\Vert Q_0)}}\D{P_1}{P_0}.
  \end{align}
  In addition, the lower bound is achieved with optimal key throughput 
  \begin{align}
    r_k=\sqrt{\frac{1}{\chi_2(Q_1\Vert Q_0)}}\left(\D{Q_1}{Q_0}-\D{P_1}{P_0}\right)^+,
  \end{align}
  where $(x)^+\eqdef\max\{x, 0\}$.
\end{theorem}
\begin{IEEEproof}
  See Appendix~\ref{sec:proof-theorem-refined}.
\end{IEEEproof}
The use of the relative entropy $D^{(n)}$ instead of variational distance and the definition of $P_{\textnormal{max}}^{(n)}$ are purely because of analytical convenience. Analyzing the secret key throughput under variational distance would require a much more sophisticated analysis of the converse in line with the resolvability analysis of~\cite{Watanabe2014Strong}. Analyzing \ac{AWGN} channels might be possible but would require different techniques in the achievability proofs as in~\cite{Zhang2019a}. Both extensions are outside the scope of this paper.

Theorem~\ref{th:refinement} nevertheless sheds light on intriguing aspects of the model. First, note that the covert capacity bounds are identical to those obtained in Theorem~\ref{thm:async-covert-capacity}. This supports the conjecture that for sub-exponential random slot selection, the choice of covertness metric has no impact on the result. Second, the characterization of the optimal key throughput shows that the condition $\D{P_1}{P_0}>\D{Q_1}{Q_0}$ is necessary to avoid the use of secret keys.

\subsection{Vanishing Covertness Constraint}
\label{sec:rmk-vanishing-covertness}
Any \emph{constant} covertness constraint $\delta$ does not change the covert capacity because $V^{(n)}\in o(1)$, as can be observed in Lemma \ref{lm:prelim-dmc}, Lemma~\ref{lm:prelim-AWGN}, and the converse arguments for both cases. If we introduce $\delta_n$ as a decaying function of $n$, we then make the following observations on the achievability for the \acp{DMC}. Recall that
    \begin{align*}
      \D{Q^N_{\alpha_n}}{Q^{\otimes N}_0} \leq \frac{e^{n\alpha_n^2\chi_2(Q_1\Vert Q_0)}-1}{L}\leq 2\delta_n^2.
    \end{align*}
    Therefore, we would choose $\alpha_n=\sqrt{\frac{\log(1+2L_n\delta_n^2)}{n\chi_2(Q_1\Vert Q_0)}}$ (we omit some small order terms to simplify discussion), and we need to consider the following two scenarios:
    \begin{enumerate}
    \item If $L_n\delta_n^2\in \omega(1)\cap \bigcap_{c>0} o(c^n)$, then the result extends and we again observe that $\log L_n$ dominates over $\log\delta_n^2$ term.
      \item If $L_n\delta_n^2\in o(1)$, then we obtain $\log(1+2L_n\delta_n^2)\approx 2L_n\delta_n^2$. In this case, we somehow obtain a different scaling of covert throughput as
        \begin{align*}
          \frac{\log M}{\sqrt{nL_n\delta_n^2}}\leq\sqrt{\frac{2}{\chi_2(Q_1\Vert Q_0)}}\D{P_1}{P_0},
        \end{align*}
        which falls back to an inferior scaling compared to the previous one, as $L_n\delta_n^2\in o(1)$.
      \end{enumerate}
      
      Unfortunately, our converse argument does not provide a fine control on how fast the covertness constraint $\delta_n$ could decay. Unlike previous works \cite{Tahmasbi2019,Zhang2019a,Wang2021}, in which the converse argument can push the operating point on the \ac{ROC} close to $(\frac{1}{2}, \frac{1}{2})$, here we only argue the impossibility of covertness (i.e., $\alpha + \beta\to0$) if the covert throughput go beyond the upper bound, which refines the idea from \cite{Bash2013,Bash2016,Dani2021Covert} to give a precise constant in front of the $\sqrt{n\log L_n}$ scaling.

\appendices

\section{Proof of Lemma~\ref{lm:DMC-ch-reliability}}
\label{sec:proof-DMC-ch-rel}

\begin{IEEEproof}
  We consider the following two decoding error events given that $\bfx_w$ is transmitted in $\ell$-th time slot:
  \begin{itemize}
  \item $(\bfx_{w},\bfy_{\intseq{n(\ell-1)+1}{n\ell}})\notin \calA_\gamma^n$;
    \item there exist codeword $\bfx_{w'}$ and time slot $k$ such that $(\bfx_{w'},\bfy_{\intseq{n(k-1)+1}{nk}})\in \calA_\gamma^n$, but either $w'\neq w$ or $k\neq \ell$.
    \end{itemize}
    The expected average probability of error proceeds as follows:
    \begin{align}
      &\E[\calC]{P_e^{(n)}}\nonumber\\
      &\leq\mathbb{E}_{\calC}\Bigg[\sum_{y^N}\sum_{\ell=1}^{L_n}\sum_{w=1}^M\frac{1}{ML_n}\WbN(y^N|X^N_{w,\ell})\indic{(\bfX_{w}, \bfy_{\intseq{n(\ell-1)+1}{n\ell}})\notin\calA_\gamma^n}\Bigg]\nonumber\\
      &\quad+\mathbb{E}_{\calC}\Biggl[\sum_{y^N}\sum_{\ell=1}^{L_n}\sum_{w=1}^M\frac{1}{ML_n}\WbN(y^N|X^N_{w,\ell})\Bigg.\nonumber\\
      &\phantom{=====}\Bigg.\times\indic{\exists w'\neq w~\text{or}~k\neq\ell~\text{s.t.}~(\bfX_{w'}, \bfy_{\intseq{n(k-1)+1}{nk}})\in\calA_\gamma^n}\Biggr]\nonumber\\
      &\labrel={eq:DMC-ch-rel-sym}\E[\calC]{\sum_{y^N}\WbN(y^N|X^N_{1,1})\indic{(\bfX_{1}, \bfy_{\intseq{1}{n}})\notin\calA_\gamma^n}}\nonumber\\
      &\quad+\mathbb{E}_{\calC}\Biggl[\sum_{y^N}\WbN(y^N|X^N_{1,1})\Bigg.\nonumber\\
      &\phantom{=}\Bigg.\times\indic{\exists w'\neq 1~\text{or}~k\neq 1~\text{s.t.}~(\bfX_{w'}, \bfy_{\intseq{n(k-1)+1}{nk}})\in\calA_\gamma^n}\Biggr]\nonumber\\
      &\labrel\leq{eq:DMC-ch-rel-uni}\E[\calC]{\sum_{y^N}\WbN(y^N|X^N_{1,1})\indic{(\bfX_{1}, \bfy_{\intseq{1}{n}})\notin\calA_\gamma^n}}\nonumber\\
      \label{eq:DMC-ch-rel-err-1}
      &\quad+\sum_{(w',k)\neq(1,1)}\mathbb{E}_{\calC}\Bigg[\sum_{y^N}\WbN(y^N|X^N_{1,1})\indic{(\bfX_{w'}, \bfy_{\intseq{n(k-1)+1}{nk}})\in\calA_\gamma^n}\Bigg],
    \end{align}
    where \eqref{eq:DMC-ch-rel-sym} follows from the symmetry and \eqref{eq:DMC-ch-rel-uni} follows from the union bound.

Following the similar techniques in \cite[Lemma 3]{Bloch2016} and \cite[Theorem 18]{Polyanskiy2010a}, we obtain
    \begin{align}\label{eq:DMC-ch-rel-err-4}
      &\E[\calC]{P_e^{(n)}}\leq\P[\Pi_{\alpha_n}^{\otimes n}\Wbn]{\log\frac{\Wbn(\bfY|\bfX)}{P_{\alpha_n}^{\otimes n}(\bfY)}\leq \gamma}+ML_ne^{-\gamma}.
    \end{align}
    We then proceed with the first term of \eqref{eq:DMC-ch-rel-err-4} as follows:
    \begin{align}\label{eq:ch-rel-non-typ-prob-1}
      &\P[\Pi_{\alpha_n}^{\otimes n}\Wbn]{\log\frac{\Wbn(\bfY|\bfX)}{P_{\alpha_n}^{\otimes n}(\bfY)}\leq \gamma}=\P[\Pi_{\alpha_n}^{\otimes n}\Wbn]{\sum_{i=1}^n\log\frac{\Wb(Y_i|X_i)}{P_{\alpha_n}(Y_i)}\leq \gamma}.
    \end{align}
    We then recall Bernstein's inequality in Lemma~\ref{lm:Bernstein}.
    \begin{lemma}[Bernstein's inequality]
      \label{lm:Bernstein}Let $U_1, U_2, \cdots, U_n\in\bbR$ be independent random variables with finite variance such that $\abs{U_i}<b$ for some $b>0$ almost surely for all $i\leq n$. Let $S=\sum_{i=1}^n(U_i-\E{U_i})$. Then, for any $t>0$,
      \begin{align}
        \P{S\geq t}\leq\exp\left(-\frac{t^2}{2\left(\sum_{i=1}^n\E{U_i^2}+\frac{bt}{3}\right)}\right).
      \end{align}
    \end{lemma}
   In order to proceed with Lemma~\ref{lm:Bernstein}, we follow key steps in \cite[Appendix D]{Arumugam2019} and obtain the first and second moment of information density $\log\frac{\Wb(Y|X)}{P_{\alpha_n}(Y)}$ as follows:
    \begin{align}\label{eq:mean-info-density-ch-rel}
      \E[\Pi_{\alpha_n}\Wb]{\log\frac{\Wb(Y|X)}{P_{\alpha_n}(Y)}}
=\alpha_n\D{P_1}{P_0}+\calO(\alpha_n^2),
    \end{align}
    
    \begin{align}\label{eq:2-mom-info-density-ch-rel}      
      &\E[\Pi_{\alpha_n}\Wb]{\log^2\frac{\Wb(Y|X)}{P_{\alpha_n}(Y)}}=\alpha_n\Lambda+\calO(\alpha_n^2),
    \end{align}
    where $\Lambda\eqdef\sum_{y}P_1(y)\log^2\left(\frac{P_1(y)}{P_0(y)}\right)$.
    
    Let $K\eqdef\max_{x, y}\abs{\log\frac{\Wb(y|x)}{P_{\alpha_n}(y)}}<\infty$. For some $\nu_1\in(0,1)$, by choosing
    \begin{align}\label{eq:ch-rel-gamma}
      \gamma&=(1-\nu_1)n\E[\Pi_{\alpha_n}\Wb]{\log\frac{\Wb(Y|X)}{P_{\alpha_n}(Y)}}\nonumber\\
      &=(1-\nu_1)n\left(\alpha_n\D{P_1}{P_0}+\calO(\alpha_n^2)\right),
    \end{align}
 we can use Lemma~\ref{lm:Bernstein} with \eqref{eq:mean-info-density-ch-rel} and \eqref{eq:2-mom-info-density-ch-rel} to analyze \eqref{eq:ch-rel-non-typ-prob-1} as follows:
    \begin{align}
      \label{eq:ch-rel-non-typ-prob-2}
      &\P[\Pi_{\alpha_n}^{\otimes n}\Wbn]{\sum_{i=1}^n\log\frac{\Wb(Y_i|X_i)}{P_{\alpha_n}(Y_i)}\leq \gamma}\nonumber\\
      &\labrel={eq:DMC-ch-rel-U}\P[\Pi_{\alpha_n}^{\otimes n}\Wbn]{n\nu_1\E{U}\leq\sum_{i=1}^n\left(\E{U}-U_i\right)}\nonumber\\
      &\leq\exp\left(-\frac{n^2\nu_1^2\left(\alpha_n\D{P_1}{P_0}+\calO(\alpha_n^2)\right)^2}{2\left(n\alpha_n\Lambda+\calO(n\alpha_n^2)+\frac{K}{3}n\nu_1\alpha_n\D{P_1}{P_0}\right)}\right)\nonumber\\
      &\leq\exp\left(-\frac{n\nu_1^2\left(\alpha_n\D{P_1}{P_0}^2+\calO(\alpha_n^2)\right)}{2\left(\Lambda+\calO(\alpha_n)+\frac{K}{3}\nu_1\D{P_1}{P_0}\right)}\right),
    \end{align}
    where \eqref{eq:DMC-ch-rel-U} follows by letting $U_i\eqdef\E[\Pi_{\alpha_n}\Wb]{\log\frac{\Wb(Y|X)}{P_{\alpha_n}(Y)}}-\log\frac{\Wb(Y_i|X_i)}{P_{\alpha_n}(Y_i)}$.
    By combining \eqref{eq:DMC-ch-rel-err-4} with \eqref{eq:ch-rel-non-typ-prob-2}, we obtain
    \begin{align*}
      &\E[\calC]{P_e^{(n)}}\leq\exp\left(-\frac{n\nu_1^2\left(\alpha_n\D{P_1}{P_0}^2+\calO(\alpha_n^2)\right)}{2\left(\Lambda+\calO(\alpha_n)+\frac{K}{3}\nu_1\D{P_1}{P_0}\right)}\right)+ML_ne^{-\gamma},
    \end{align*}
     then by choosing $\log M=(1-\delta_1)(1-\nu_1)n\left(\alpha_n\D{P_1}{P_0}+\calO(\alpha_n^2)\right)$ for some $\delta_1\in(0,1)$ and from the choice of $\alpha_n$, there exists some $\xi_1>0$ such that for $n$ large enough
    \begin{align}
      \E[\calC]{P_e^{(n)}}\leq\exp\left(-\xi_1\sqrt{n\log L_n}\right).
    \end{align}
\end{IEEEproof}

\section{Proof of Lemma~\ref{lm:DMC-ch-rsl}}
\label{sec:proof-DMC-ch-rsl}

\begin{IEEEproof}
  Note first that given any randomly generated codebook $\calC$, from \eqref{eq:Q-N-T-hat} and \eqref{eq:DMC-Q-alpha-N}, we identify that,
  \begin{align}\label{eq:Q-code-simp}
    \V{\QcodeN}{Q_{\alpha_n}^N}&=\V{\E[T]{\widehat{Q}_{\alpha_n, T}^N}}{\E[T]{Q_{\alpha_n, T}^N}}\nonumber\\
                               &\labrel\leq{eq:DMC-ch-rsl-mono}\E[T]{\V{\widehat{Q}^N_{Z, T}}{Q_{\alpha_n, T}^{N}}}\nonumber\\
    &\labrel={eq:DMC-ch-rsl-symm}\V{\widehat{Q}^N_{Z, 1}}{Q_{\alpha_n, 1}^N}=\V{\Qcoden}{Q_{\alpha_n}^{\otimes n}},
  \end{align}
  where \eqref{eq:DMC-ch-rsl-mono} follows from the fact that $\V{\cdot}{\cdot}$ belongs to the family of $f$-divergence and the monotonicity of $f$-divergence, \eqref{eq:DMC-ch-rsl-symm} follows from the fact that $T$ is uniformly distributed and the symmetry, and $\Qcoden(\bfz)\eqdef\frac{1}{M}\sum_{w=1}^M\Wwn(\bfz|\bfx_w)$. Therefore, we can now focus on the transmission that occurs during a time slot known to every party (synchronous regime). Note also that $Q_{\alpha_n}^{\otimes n}(\bfz)=\E[\calC]{\smash{\Qcoden}(\bfz)}$.
  
  To analyze $\E[\calC]{\smash{\V{\smash{\Qcoden}}{Q_{\alpha_n}^{\otimes n}}}}$, we follow techniques similar to \cite[Theorem VII.1]{Cuff2013} and \cite[Lemma 5]{Bloch2016} and obtain
  \begin{align}
    \label{eq:DMC-ch-rsl-V-2}
    &\E[\calC]{\V{\Qcoden}{Q_{\alpha_n}^{\otimes n}}}\leq\P[\Pi_{\alpha_n}^{\otimes n}\Wwn]{\log\frac{\Wwn(\bfZ|\bfX)}{Q_{\alpha_n}^{\otimes n}(\bfZ)}\geq\tau}+\frac{1}{2}\sqrt{\frac{e^\tau}{M}}.
  \end{align}
  We then proceed with the first term of \eqref{eq:DMC-ch-rsl-V-2} as follows:
  \begin{align}\label{eq:DMC-ch-rsl-V-non-typ}
    &\P[\Pi_{\alpha_n}^{\otimes n}\Wwn]{\log\frac{\Wwn(\bfZ|\bfX)}{Q_{\alpha_n}^{\otimes n}(\bfZ)}\geq\tau}=\P[\Pi_{\alpha_n}^{\otimes n}\Wwn]{\sum_{i=1}^n\log\frac{\Ww(Z_i|X_i)}{Q_{\alpha_n}(Z_i)}\geq\tau}.
  \end{align}
  Similar to~\eqref{eq:ch-rel-non-typ-prob-2}, upon recognizing
  \begin{align}
    \E[\Pi_{\alpha_n}\Ww]{\log\frac{\Ww(Z|X)}{Q_{\alpha_n}(Z)}}&=\alpha_n\D{Q_1}{Q_0}+\calO(\alpha_n^2),\\
    \E[\Pi_{\alpha_n}\Ww]{\log^2\frac{\Ww(Z|X)}{Q_{\alpha_n}(Z)}}&=\alpha_n\Gamma+\calO(\alpha_n^2),
  \end{align}
  where $\Gamma\eqdef\sum_{z}Q_1(z)\log^2\frac{Q_1(z)}{Q_0(z)}$ and $\overline{K}\eqdef\max_{x,z}\abs{\log\frac{W_{Z|X}(z|x)}{Q_{\alpha_n}(z)}}$, by choosing
  \begin{align}
    \tau=n(1+\nu_2)\left(\alpha_n\D{Q_1}{Q_0}+\calO(\alpha_n^2)\right)
  \end{align}
  for some $\nu_2\in(0,1)$, we can use Lemma~\ref{lm:Bernstein} to analyze \eqref{eq:DMC-ch-rsl-V-non-typ} as follows
  \begin{align}
    &\P[\Pi_{\alpha_n}^{\otimes n}\Wwn]{\sum_{i=1}^n\log\frac{\Ww(Z_i|X_i)}{Q_{\alpha_n}(Z_i)}\geq\tau}\nonumber\\
    &\labrel={eq:DMC-ch-rsl-rv}\P[\Pi_{\alpha_n}^{\otimes n}\Wwn]{\sum_{i=1}^n\left(U_i-\E{U}\right)\geq n\nu_2\E{U}}\nonumber\\
    &\leq\exp\left(-\frac{n^2\nu_2^2\left(\alpha_n\D{Q_1}{Q_0}+\calO(\alpha_n^2)\right)^2}{2\left(n\alpha_n\Gamma+\calO(n\alpha_n^2)+\frac{\overline{K}}{3}n\nu_2\alpha_n\D{Q_1}{Q_0}\right)}\right)\nonumber\\
   \label{eq:DMC-ch-rsl-V-non-typ-2} &\leq\exp\left(-\frac{n\nu_2^2\left(\alpha_n\D{Q_1}{Q_0}+\calO(\alpha_n^2)\right)}{2\left(\Gamma+\calO(\alpha_n)+\frac{\overline{K}}{3}\nu_2\D{Q_1}{Q_0}\right)}\right),
  \end{align}
  where \eqref{eq:DMC-ch-rsl-rv} follows by letting $U_i\eqdef\log\frac{\Ww(Z_i|X_i)}{Q_{\alpha_n}(Z_i)}-\E[\Pi_{\alpha_n}\Ww]{\log\frac{\Ww(Z|X)}{Q_{\alpha_n}(Z)}}$. By combining \eqref{eq:DMC-ch-rsl-V-2} with \eqref{eq:DMC-ch-rsl-V-non-typ-2}, we obtain
  \begin{align*}
    &\E[\calC]{\V{\Qcoden}{Q_{\alpha_n}^{\otimes n}}}\leq\exp\left(-\frac{n\nu_2^2\left(\alpha_n\D{Q_1}{Q_0}+\calO(\alpha_n^2)\right)}{2\left(\Gamma+\calO(\alpha_n)+\frac{\overline{K}}{3}\nu_2\D{Q_1}{Q_0}\right)}\right)+\frac{1}{2}\sqrt{\frac{e^\tau}{M}},
  \end{align*}
  then by choosing $\log M=(1+\delta_2)(1+\nu_2)n\left(\alpha_n\D{Q_1}{Q_0}+\calO(\alpha_n^2)\right)$ for some $\delta_2\in(0,1)$ and from the choice of $\alpha_n$, there exists some $\xi_2>0$ such that for $n$ large enough
  \begin{align}\label{eq:DMC-ch-rsl-V-3}
    \E[\calC]{\V{\Qcoden}{Q_{\alpha_n}^{\otimes n}}}\leq\exp\left(-\xi_2\sqrt{n\log L_n}\right).
  \end{align}
  Consequently, combining~\eqref{eq:Q-code-simp} with~\eqref{eq:DMC-ch-rsl-V-3}, we obtain
  \begin{align}
    \E[\calC]{\V{\QcodeN}{Q_{\alpha_n}^{N}}}\leq\exp\left(-\xi_2\sqrt{n\log L_n}\right).
  \end{align}
\end{IEEEproof} 

\section{Proof of Lemma~\ref{lm:DMC-conv-min-weight}}
\label{sec:proof-DMC-conv-min-weight}
\begin{IEEEproof}
We first construct a test as follows:
\begin{align*}
  T(\bfZ_\ell)\eqdef\frac{1}{n}\sum_{i=1}^nA(Z_{\ell,i}),
\end{align*}
where $\ell\in[1, L_n]$, $Z_{\ell,i}\eqdef Z_{n(\ell-1)+i}$, and $A(z)\eqdef\frac{Q_1(z)-Q_0(z)}{Q_0(z)}$. Then
\begin{align}
  \alpha&\eqdef\P[H_0]{\max_{\ell\in\intseq{1}{L_n}}T(\bfZ_\ell)> \tau}\leq\sum_{\ell=1}^{L_n}\P[H_0]{T(\bfZ_\ell)> \tau}\leq L_n\P[H_0]{T(\bfZ_\ell)> \tau}.
\end{align}

To apply Lemma~\ref{lm:Bernstein}, we first let $C\eqdef\max_{z\in\calZ}A(z)$. Then $C<\infty$, since $Q_1\ll Q_0$ and $\card{\calZ}<\infty$. Therefore,
\begin{align}
  \P[H_0]{T(\bfZ)>\tau}&=\P[H_0]{\frac{1}{n}\sum_{i=1}^nA(z_i)>\tau}\leq\exp\left(-\frac{\frac{1}{2}n^2\tau^2}{n\chi_2+\frac{1}{3}nC\tau}\right).
\end{align}
There exists a choice of $\tau\eqdef\epsilon\sqrt{\frac{2\chi_2\log L_n}{n}}$ with $\epsilon>1$ such that
\begin{align}
  \exp\left(-\frac{\frac{1}{2}n^2\tau^2}{n\chi_2+\frac{1}{3}nC\tau}\right)&\leq\exp\left(-\frac{n\tau^2}{2\chi_2}+\frac{n\tau^3C}{6\chi_2^2}\right)\nonumber\\
  &=\exp(-(\epsilon-o(1))\log L_n).
\end{align}
This implies that
\begin{align}
  \alpha\leq\exp(-(\epsilon-1-o(1))\log L_n)
\end{align}
and can be made arbitrarily small with increasing $n$, as $\log L_n\in o(n)\cap \omega(1)$.
Now we let $w_m\eqdef\wt{\bfx_m}$ for every $\bfx_m\in\calC$ and assume that the transmission occurs in slot $\ell$ and $w_*\eqdef\min_mw_m\geq\epsilon\sqrt{\frac{2n\log L_n}{\chi_2}}$.
For the probability of missed-detection, we have for every $m\in\intseq{1}{M}$,
\begin{align}
&\beta_m\leq\P[H_1]{\frac{1}{n}\sum_{i=1}^nA(Z_{\ell, i})\leq\tau}\\
&=\mathbb{P}_{H_1}\bigg(\sum_{i=1}^nA(Z_{_{\ell,i}})-\left(n\mu_0+w_m(\mu_1-\mu_0)\right)\leq n\tau-\left(n\mu_0+w_m(\mu_1-\mu_0)\right)\bigg)\\
&\leq\mathbb{P}_{H_1}\Bigg(\abs{\sum_{i=1}^nA(Z_{_{\ell,i}})-\left(n\mu_0+w_m(\mu_1-\mu_0)\right)}\geq\left(n\mu_0+w_m(\mu_1-\mu_0)\right)-n\tau\Bigg)\\
&\labrel\leq{eq:conv-DMC-cheby}\frac{n\sigma_0^2+w_m(\sigma_1^2-\sigma_0^2)}{\left(w_m\chi_2-\epsilon\sqrt{2n\chi_2\log L_n}\right)^2},
\end{align}
where we have let $\mu_0\eqdef\E[\Qinn]{A(Z)}$, $\sigma_0^2\eqdef\Var[\Qinn]{A(Z)}$, $\mu_1\eqdef\E[\Qinf]{A(Z)}$, and $\sigma_1^2\eqdef\Var[\Qinf]{A(Z)}$, and \eqref{eq:conv-DMC-cheby} follows by Chebyshev's inequality. If $\sigma_1^2\leq\sigma_0^2$,
\begin{align*}
  \frac{n\sigma_0^2+w_m(\sigma_1^2-\sigma_0^2)}{\left(w_m\chi_2-\epsilon\sqrt{2n\chi_2\log L_n}\right)^2}\leq\frac{n\sigma_0^2}{\left(w_*\chi_2-\epsilon\sqrt{2n\chi_2\log L_n}\right)^2}.
\end{align*}
On the other hand, if $\sigma_1^2>\sigma_0^2$, then we proceed as follows:
\begin{inparaenum}[1)]
  \item if $w_m>\frac{\sigma_1}{\sigma_0}w_*$,
    \begin{align*}
      \frac{n\sigma_0^2+w_m(\sigma_1^2-\sigma_0^2)}{\left(w_m\chi_2-\epsilon\sqrt{2n\chi_2\log L_n}\right)^2}&\leq\frac{n\sigma_1^2}{\left(\frac{\sigma_1}{\sigma_0}w_*\chi_2-\epsilon\sqrt{2n\chi_2\log L_n}\right)^2}\\
                                                                                                           &\leq\frac{n\sigma_0^2}{\left(w_*\chi_2-\frac{\sigma_0}{\sigma_1}\epsilon\sqrt{2n\chi_2\log L_n}\right)^2}\\
      &\leq\frac{n\sigma_0^2}{\left(w_*\chi_2-\epsilon\sqrt{2n\chi_2\log L_n}\right)^2};
    \end{align*}
    \item if $\frac{\sigma_1}{\sigma_0}w_*\geq w_m\geq w_*$,
      \begin{align*}
        \frac{n\sigma_0^2+w_m(\sigma_1^2-\sigma_0^2)}{\left(w_m\chi_2-\epsilon\sqrt{2n\chi_2\log L_n}\right)^2}\leq\frac{n\sigma_0^2+\frac{\sigma_1}{\sigma_0}w_*(\sigma_1^2-\sigma_0^2)}{\left(w_*\chi_2-\epsilon\sqrt{2n\chi_2\log L_n}\right)^2}.
      \end{align*}
    \end{inparaenum}
    Therefore,
    \begin{align}
      \beta=\frac{1}{M}\sum_m\beta_m\leq\frac{n\sigma_0^2+\frac{\sigma_1}{\sigma_0}w_*(\sigma_1^2-\sigma_0^2)}{\left(w_*\chi_2-\epsilon\sqrt{2n\chi_2\log L_n}\right)^2}.
    \end{align}
If $w_*\geq\epsilon\sqrt{\frac{2n\log L_n}{\chi_2}}$ for any $\epsilon>1$, then there exists some $\tau$ of Willie such that $\alpha+\beta\to0$ meaning it cannot be covert. 
\end{IEEEproof}

\section{Proof of Lemma~\ref{lm:AWGN-ch-rel}}
\label{sec:proof-AWGN-ch-rel}
\begin{IEEEproof}
  We first recall the following lemma that plays the same role as \eqref{eq:DMC-ch-rel-err-4} to analyze the channel reliability with the decoding region $\overline{\calA}_\gamma^n$. 
  \begin{lemma}[Modified from {\cite[Lemma 3]{Bloch2016}, \cite[Lemma 1]{Tahmasbi2019}, and \cite[Lemma 14]{Wang2021}}]\label{lm:AWGN-ch-rel-info}
    For any $\gamma>0$,
    \begin{align}\label{eq:AWGN-ch-rel-err-1}
      \E[\calC]{P_e^{(n)}}&\leq ML_ne^{-\gamma}\E[P_{\rho_n}^{\otimes n}]{\frac{P_{\rho_n}^{\otimes n}(\bfY)}{\Pinn^{\otimes n}(\bfY)}}+\P[\Pi_{\rho_n}^{\otimes n}\Wbn]{\log\frac{\Wbn(\bfY|\bfX)}{\Pinn^{\otimes n}(\bfY)}\leq\gamma}.
    \end{align}
  \end{lemma}
  Note that the main difference from \eqref{eq:DMC-ch-rel-err-4} is that $\overline{\calA}_\gamma^n$ is defined \ac{wrt} $\log\frac{\Wbn(\bfY|\bfX)}{\Pinn^{\otimes n}(\bfY)}$ and it consequently introduces an additional factor $\E[P_{\rho_n}^{\otimes n}]{\frac{P_{\rho_n}^{\otimes n}(\bfY)}{\Pinn^{\otimes n}(\bfY)}}$ to account for the change of measure. This modification is to simplify the analysis on information density into a Gaussian random variable while, as we shall see next, the additional penalty term is properly bounded. Note that \cite{wangGaussianCovertCommunication2019} directly analyzes $\log\frac{\Wbn(\bfY|\bfX)}{P_{\rho_n}^{\otimes n}(\bfY)}$, and the key step therein is to develop the lower bound \cite[(43)]{wangGaussianCovertCommunication2019} for information density. However, as we shall see in Appendix~\ref{sec:proof-AWGN-ch-rsl} for Lemma~\ref{lm:AWGN-ch-rsl}, techniques in~\cite{wangGaussianCovertCommunication2019} does not extend since it requires an upper bound on the information density. Our techniques are more suitable to proceed with both cases since we do not require such upper and lower bounds on the information density.

  For the penalty term, 
  \begin{align}\label{eq:AWGN-ch-rel-penalty}
    &\E[P_{\rho_n}^{\otimes n}]{\frac{P_{\rho_n}^{\otimes n}(\bfY)}{\Pinn^{\otimes n}(\bfY)}}\nonumber\\
    &=\left(\E[P_{\rho_n}]{\frac{P_{\rho_n}(Y)}{\Pinn(Y)}}\right)^n=\left(\cosh\left(\frac{\rho_n}{\sigma_b^2}\right)\right)^n\nonumber\\
                                                                                            &\labrel\leq{eq:AWGN-ch-rel-cosh}\exp\left(n\frac{\rho_n^2}{2\sigma_b^4}\right)\nonumber\\
    &\labrel={eq:AWGN-ch-rel-penalty-order}\exp\left(\frac{\sigma_w^4}{\sigma_b^4}\log\left(L_n(2\delta^2-4n^{-\frac{1}{2}})\right)\right),
  \end{align}
  where \eqref{eq:AWGN-ch-rel-cosh} follows from $\cosh(x)\leq\exp(\frac{x^2}{2})$ for $x\in\bbR$, and \eqref{eq:AWGN-ch-rel-penalty-order} follows from \eqref{eq:AWGN-achv-rho}.
  For the second term of~\eqref{eq:AWGN-ch-rel-err-1}, we start by noting that
  \begin{align*}
    \log\frac{\Wbn(\bfY|\bfX)}{\Pinn^{\otimes n}(\bfY)}&=\sum_{i=1}^n\log\frac{\Wb(Y_i|X_i)}{\Pinn(Y_i)}=\sum_{i=1}^n\left(\frac{X_iY_i}{\sigma_b^2}-\frac{X_i^2}{2\sigma_b^2}\right).
  \end{align*}
  Since our \ac{BPSK} codewords require $X_i\in\{-a_n, a_n\}$ for every $i\in\intseq{1}{n}$, we therefore identify that
  \begin{align*}
    \log\frac{\Wb(Y_i|X_i)}{\Pinn(Y_i)}\sim\calN\left(\frac{\rho_n}{2\sigma_b^2}, \frac{\rho_n}{\sigma_b^2}\right).
  \end{align*}
  \begin{lemma}[Chernoff Bound for Sum of Independent Gaussian Random Variables]
    \label{lm:AWGN-chernoff}
    Let $U_1, U_2, \cdots, U_n\in\bbR$ be zero-mean and independent Gaussian random variables distributed according to $U_i\sim\calN(0, \sigma_i^2)$. Let $S=\sum_{i=1}^nU_i$. Then, for any $t>0$,
    \begin{align}
      \P{S\geq t}\leq\exp\left(-\frac{t^2}{2\sum_{i=1}^n\sigma_i^2}\right).
    \end{align}
  \end{lemma}
  Therefore, by setting $\gamma=(1-\nu_1)n\frac{\rho_n}{2\sigma_b^2}$ for some $\nu\in(0,1)$, Lemma~\ref{lm:AWGN-chernoff} implies that
  \begin{align}\label{eq:AWGN-ch-rel-non-typ}
    &\P[\Pi_{\rho_n}^{\otimes n}\Wbn]{\log\frac{\Wbn(\bfY|\bfX)}{\Pinn^{\otimes n}(\bfY)}\leq\gamma}\nonumber\\
    &=\P[\Pi_{\rho_n}^{\otimes n}\Wbn]{\sum_{i=1}^n\log\frac{\Wb(Y_i|X_i)}{\Pinn(Y_i)}\leq\gamma}\nonumber\\
    &=\sum_{\bfx\in\{-a_n,a_n\}^n}\Pi_{\rho_n}^{\otimes n}(\bfx)\P[W_{\bfY|\bfX=\bfx}^{\otimes n}]{\sum_{i=1}^n\log\frac{\Wb(Y_i|x_i)}{\Pinn(Y_i)}\leq\gamma}\nonumber\\
    &\labrel={eq:AWGN-ch-rel-Gau}\sum_{\bfx\in\{-a_n,a_n\}^n}\Pi_{\rho_n}^{\otimes n}(\bfx)\P[W_{\bfY|\bfX=\bfx}^{\otimes n}]{\sum_{i=1}^nU_i\geq n\nu_1\frac{\rho_n}{2\sigma_b^2}}\nonumber\\
    &\leq\sum_{\bfx\in\{-a_n,a_n\}^n}\Pi_{\rho_n}^{\otimes n}(\bfx)\exp\left(-\frac{n\nu_1^2\rho_n}{8\sigma_b^2}\right)\nonumber\\
    &=\exp\left(-\frac{n\nu_1^2\rho_n}{8\sigma_b^2}\right),
  \end{align}
  where \eqref{eq:AWGN-ch-rel-Gau} follows by defining $U_i=\frac{\rho_n}{2\sigma_b^2}-\log\frac{\Wb(Y_i|x_i)}{\Pinn(Y_i)}$. Then, by combining \eqref{eq:AWGN-ch-rel-penalty} and \eqref{eq:AWGN-ch-rel-non-typ} with \eqref{eq:AWGN-ch-rel-err-1}, we obtain
  \begin{align}
    &\E[\calC]{P_e^{(n)}}\nonumber\\
    &\leq Me^{-\gamma}\exp\left(\frac{\sigma_w^4}{\sigma_b^4}\log\left(L_n(2\delta^2-4n^{-\frac{1}{2}})\right) +\log L_n\right)+\exp\left(-\frac{n\nu_1^2\rho_n}{8\sigma_b^2}\right)\nonumber\\
    &=M\exp\bigg(\frac{\sigma_w^4}{\sigma_b^4}\log\left(L_n(2\delta^2-4n^{-\frac{1}{2}})\right) +\log L_n-(1-\nu_1)\frac{n\rho_n}{2\sigma_b^2}\bigg)+\exp\left(-\frac{n\nu_1^2\rho_n}{8\sigma_b^2}\right),
  \end{align}
  then by choosing $\log M=(1-\delta_1)(1-\nu_1)\frac{n\rho_n}{2\sigma_b^2}$ for some $\delta_1\in(0,1)$, from \eqref{eq:AWGN-achv-rho} and $\log L_n\in o(n)$, we obtain
  \begin{align}
    \E[\calC]{P_e^{(n)}}\leq\exp\left(-\theta_1\sqrt{n\log L_n}\right)
  \end{align}
  for some $\theta_1>0$ and $n$ large enough.
\end{IEEEproof}
\section{Proof of Lemma~\ref{lm:AWGN-ch-rsl}}
\label{sec:proof-AWGN-ch-rsl}

\begin{IEEEproof}
  Similar to the proof of Lemma~\ref{lm:DMC-ch-rsl}, we first identify that given any randomly generated codebook $\calC$,
  \begin{align}
    \label{eq:AWGN-ch-rsl-convex}
    \V{\QcodeN}{Q^N_{\rho_n}}\leq\V{\Qcoden}{Q_{\rho_n}^{\otimes n}},
  \end{align}
  where $\Qcoden(\bfz)\eqdef\frac{1}{M}\sum_{w=1}^M\Wwn(\bfz|\bfx_w)$. Note also that $Q_{\rho_n}^{\otimes n}=\E[\calC]{\Qcoden}$. To analyze $\E[\calC]{\V{\Qcoden}{Q_{\rho_n}^{\otimes n}}}$, we recall the following lemma:
  \begin{lemma}[{\cite[Lemma 5]{Bloch2016}}]
    \label{lm:AWGN-ch-rsl-info}
    For any $\tau>0$,
    \begin{align}\label{eq:AWGN-ch-rsl-V-1}
      &\E[\calC]{\V{\Qcoden}{Q_{\rho_n}^{\otimes n}}}\leq\P[\Pi_{\rho_n}^{\otimes n}\Wwn]{\log\frac{\Wwn(\bfZ|\bfX)}{\Qinn^{\otimes n}(\bfZ)}\geq\tau}+\frac{1}{2}\sqrt{\frac{e^\tau}{M}}.
    \end{align}
  \end{lemma}
  For the first term of \eqref{eq:AWGN-ch-rsl-V-1}, note that
  \begin{align*}
    \log\frac{\Wwn(\bfZ|\bfX)}{\Qinn^{\otimes n}(\bfZ)}&=\sum_{i=1}^n\log\frac{\Ww(Z_i|X_i)}{\Qinn(Z_i)}=\sum_{i=1}^n\left(\frac{X_iZ_i}{\sigma_w^2}-\frac{X_i^2}{2\sigma_w^2}\right).
  \end{align*}
  Since our \ac{BPSK} codewords require $X_i\in\{-a_n,a_n\}$ for every $i\in\intseq{1}{n}$, we therefore identify that
  \begin{align*}
    \log\frac{\Ww(Z_i|X_i)}{\Qinn(Z_i)}\sim\calN\left(\frac{\rho_n}{2\sigma_w^2},\frac{\rho_n}{\sigma_w^2}\right).
  \end{align*}
  Therefore, by setting $\tau=(1+\nu_2)n\frac{\rho_n}{2\sigma_w^2}$ for some $\nu\in(0,1)$, Lemma~\ref{lm:AWGN-chernoff} implies that
  \begin{align}
    \label{eq:AWGN-ch-rsl-non-typ}
    &\P[\Pi_{\rho_n}^{\otimes n}\Wwn]{\log\frac{\Wwn(\bfZ|\bfX)}{\Qinn^{\otimes n}(\bfZ)}\geq\tau}\nonumber\\
    &=\P[\Pi_{\rho_n}^{\otimes n}\Wwn]{\sum_{i=1}^n\log\frac{\Ww(Z_i|X_i)}{\Qinn(Z_i)}\geq\tau}\nonumber\\
    &=\sum_{\bfx\in\{-a_n, a_n\}^n}\Pi_{\rho_n}^{\otimes n}(\bfx)\P[W_{\bfZ|\bfX=\bfx}]{\sum_{i=1}^n\log\frac{\Ww(Z_i|x_i)}{\Qinn(Z_i)}\geq\tau}\nonumber\\
    &\labrel={eq:AWGN-ch-rsl-rv}\sum_{\bfx\in\{-a_n, a_n\}^n}\Pi_{\rho_n}^{\otimes n}(\bfx)\P[W_{\bfZ|\bfX=\bfx}]{\sum_{i=1}^nU_i\geq n\nu_2\frac{\rho_n}{2\sigma_w^2}}\nonumber\\
    &\leq\sum_{\bfx\in\{-a_n, a_n\}^n}\Pi_{\rho_n}^{\otimes n}(\bfx)\exp\left(-\frac{n\nu_2^2\rho_n}{8\sigma_w^2}\right)\nonumber\\
    &=\exp\left(-\frac{n\nu_2^2\rho_n}{8\sigma_w^2}\right),
  \end{align}
  where \eqref{eq:AWGN-ch-rsl-rv} follows by defining $U_i=\log\frac{\Ww(Z_i|x_i)}{\Qinn(Z_i)}-\frac{\rho_n}{2\sigma_w^2}$. Then, by combining \eqref{eq:AWGN-ch-rsl-non-typ} with \eqref{eq:AWGN-ch-rsl-V-1}, we obtain
  \begin{align}
    \E[\calC]{\V{\Qcoden}{Q_{\rho_n}^{\otimes n}}}\leq\exp\left(-\frac{n\nu_2^2\rho_n}{8\sigma_w^2}\right)+\frac{1}{2}\sqrt{\frac{e^\tau}{M}}.
  \end{align}
  Choosing $\log M=(1+\delta_2)(1+\nu_2)n\frac{\rho_n}{2\sigma_w^2}$ for some $\delta_2\in(0,1)$, from \eqref{eq:AWGN-achv-rho}, we obtain
  \begin{align}\label{eq:AWGN-ch-rsl-V-2}
    \E[\calC]{\V{\Qcoden}{Q_{\rho_n}^{\otimes n}}}\leq\exp\left(-\theta_2\sqrt{n\log L_n}\right),
  \end{align}
  for some $\theta_2>0$ and $n$ large enough. Consequently, combining \eqref{eq:AWGN-ch-rsl-convex} with~\eqref{eq:AWGN-ch-rsl-V-2}, we obtain
  \begin{align}
    \E[\calC]{\V{\QcodeN}{Q_{\rho_n}^{N}}}\leq\exp\left(-\theta_2\sqrt{n\log L_n}\right).
  \end{align}
\end{IEEEproof}

\section{Proof of Lemma~\ref{lm:AWGN-conv-V-min-power}}
\label{sec:proof-AWGN-conv-V-min-power}
\begin{IEEEproof}
  We first construct a power test as follows:
\begin{align*}
  T(\bfZ_\ell)\eqdef\frac{1}{n}\sum_{i=1}^n\abs{Z_{\ell, i}}^2-\sigma_w^2,
\end{align*}
where $\ell\in[1, L_n]$, $Z_{\ell,i}\eqdef Z_{n(\ell-1)+i}$. Then
\begin{align}
  \alpha&\eqdef\P[H_0]{\max_{\ell\in[1,L]}T(\bfZ_\ell)> \tau}\leq\sum_{\ell=1}^{L_n}\P[H_0]{T(\bfZ_\ell)> \tau}\leq L_n\P[H_0]{T(\bfZ_\ell)> \tau}.
\end{align}
\begin{lemma}[Modified from~{\cite[Corollary 2.11]{boucheronConcentrationInequalitiesNonasymptotic2013}}]
  \label{lm:chi-2-concentrate}
  Let $Z_i\sim\calN(0, \sigma^2)$ for every $i\in\intseq{1}{n}$. Then for $0<c<n\sigma^2$,
  \begin{align*}
    \P{\frac{1}{n}\sum_{i=1}^n\left(\abs{Z_i}^2-\sigma^2\right)>c}\leq\exp\left(-\frac{nc^2}{4\sigma^4+4\sigma^2c}\right).
  \end{align*}
\end{lemma}

By applying Lemma~\ref{lm:chi-2-concentrate}, there exists a choice of $\tau\eqdef\epsilon\sqrt{\frac{4\sigma_w^4\log L_n}{n}}$ with $\epsilon>1$ such that
\begin{align*}
  \exp\left(-\frac{n\tau^2}{4\sigma_w^4+4\sigma_w^2\tau}\right)&\leq\exp\left(-\frac{n\tau^2}{4\sigma_w^4}+\frac{n\tau^3}{4\sigma_w^6}\right)\\
  &=\exp(-(\epsilon-o(1))\log L_n).
\end{align*}
It therefore implies that $\alpha\leq\exp(-(\epsilon-1-o(1))\log L_n)$ and can be made arbitrarily small with increasing $n$, as $\log L_n\in o(n)\cap\omega(1)$. Now we let $P_m\eqdef\sum_{i=1}^n\abs{x_{\ell,i}^{(m)}}^2$ and assume that the transmission occurs in slot $\ell$ and $P_*\eqdef\min_mP_m\geq\epsilon\sqrt{4n\sigma_w^4\log L_n}$.  For the probability of missed-detection, we have for every $\bfx^{(m)}\in\calC$,
\begin{align*}
  \beta_m&\leq\P[H_1]{T(\bfZ_\ell)\leq\tau}\\
         &\leq\P[H_1]{\sum_{i=1}^n\abs{Z_{\ell,i}}^2-(n\sigma_w^2+P_m)\leq n\tau-P_m}\\
  &\leq\P[H_1]{\abs{\sum_{i=1}^n\abs{Z_{\ell,i}}^2-(n\sigma_w^2+P_m)}> P_m-n\tau}\\
         &\leq\frac{2n\sigma_w^4+4\sigma_w^2P_m}{(P_m-\epsilon\sqrt{4n\sigma_w^4\log L_n})^2}\\
  &\labrel\leq{eq:P-min-cheby}\frac{2n\sigma_w^4+4\sigma_w^2P_*}{(P_*-\epsilon\sqrt{4n\sigma_w^4\log L_n})^2},
\end{align*}
where \eqref{eq:P-min-cheby} follows since $\frac{aP_m}{(cP_m-d)^2}\leq\frac{aP_*}{(cP_*-d)^2}$ for $cP_*>d$ and $a, c, d>0$. Therefore,
\begin{align}
  \beta=\frac{1}{M}\sum_{m=1}^M\beta_m\leq\frac{2n\sigma_w^4+4\sigma_w^2P_*}{(P_*-\epsilon\sqrt{4n\sigma_w^4\log L_n})^2}.
\end{align}
If $P_*\geq\epsilon\sqrt{4n\sigma_w^4\log L}$ for any $\epsilon>1$, then there exists some $\tau$ of Willie such that $\alpha+\beta\to 0$.
\end{IEEEproof}

\section{Proof of Theorem~\ref{th:refinement}}
\label{sec:proof-theorem-refined}

\subsection{Sketch of Achievability Proof}
\label{sec:achievability-refined}

For brevity, we only highlight key differences with the proof of Theorem~\ref{thm:async-covert-capacity}. Note that Lemma~\ref{lm:prelim-dmc}, which is central to the analysis, is already established in terms of relative entropy and can be directly used. The channel reliability analysis also follows with minor modifications as in~\cite{Zhang2019a} to handle the modified probability of error. Using the reverse Pinsker's inequality~\cite[Eq.~(323)]{Sason2016} allows one to reuse the soft-covering analysis. One merely needs to replace $\log M$ by $\log M + \log K$ since the secret key affords extra randomization in Lemma~\ref{lm:DMC-ch-rsl} and to replace the triangle inequality in~(\ref{eq:triangleq}) as done in~\cite[Eq.~(79)]{Bloch2016} for the result to follow. 

\subsection{Sketch of Converse Proof}
\label{sec:converse-refined}

The converse proof also reuses some of the elements of proof of Theorem~\ref{thm:async-covert-capacity} but requires more care. We first establish an upper bound on $\log M$. Thanks to Pinsker's inequality (we have used $\V{\smash{\QcodeN}}{\Qinn^{\otimes N}}\leq\sqrt{\D{\smash{\QcodeN}}{\Qinn^{\otimes N}}}\leq\sqrt{\delta}$ for simplicity), we note that Lemma~\ref{lm:DMC-conv-min-weight} still applies and one can assert that there exists a sub-codebook $\calC^{(\ell)}$ containing a non-vanishing fraction $\frac{1-\sqrt{\delta}}{2}$ of the original code and with maximum weight $\sqrt{\frac{2n\log L_n}{\chi_2(Q_1\Vert Q_0)}}$.  
Following~\cite{Zhang2019a}, there also exists a sub-codebook $\calC_s^{(\ell)}$ of codewords indexed by key $s$ for which the maximum weight is also $\sqrt{\frac{2n\log L_n}{\chi_2(Q_1\Vert Q_0)}}$ and for which the average probability of error is still small by Markov's inequality and the definition of $P_{\textnormal{max}}^{(n)}$. Consequently, we obtain for any $\epsilon>0$
\begin{align}
  \frac{\log M}{\sqrt{n\log L_n}} \leq \sqrt{\frac{2}{\chi_2(Q_1\Vert Q_0)}}  (\D{P_1}{P_0}+\epsilon)\label{eq:step-in-converse},
\end{align}
as long as $n$ is large enough.

We now establish a lower bound on $\log MK$. Denote by $\mu_n$ the average weight of a codeword in the code. First note that, following the standard converse for reliability, we obtain for any $\epsilon>0$ that
\begin{align}
  \frac{\log M}{\sqrt{n\log L_n}}\leq \frac{n\mu_n}{\sqrt{n \log L_n}}(\D{P_1}{P_0} +\epsilon)
\end{align}
for $n$ large enough. We next invoke an astute argument from~\cite{Bounhar2024Whispering}. Assume that a non-zero throughput $r$ is achievable. Because of the upper bound in~(\ref{eq:step-in-converse}), there must exist $\beta\in(0;1]$ such that $r=\beta \sqrt{\frac{2}{\chi_2(Q_1\Vert Q_0)}}\D{P_1}{P_0}$ and we obtain for $n$ large enough
\begin{align}
  \frac{n\mu_n}{\sqrt{n \log L_n}}\geq \beta\sqrt{\frac{2}{\chi_2(Q_1\Vert Q_0)}} -\epsilon',
\end{align}
where $\epsilon'=O(\epsilon)$.

Upon denoting by $T$ the random variable selecting the slot, we have
\begin{align}
  \log MK
          &= H(WS)= H(WST) - \log L_n\nonumber\\
          &\geq \avgI{WST;Z^N} - \log L_n \nonumber\\
  &\geq \avgI{X^N;Z^N} - \log L_n,
\end{align}
where we have used the fact that $T$ is independent of $(W,K)$ and uniformly distributed over $L_n$ slots, and the fact that $WST-X^N-Z^N$ forms a Markov Chain. Next following standard steps as in~\cite{Hou2014}, we lower bound $I(X^N;Z^N)$ as
\begin{align}
  &\avgI{X^N;Z^N}  \nonumber\\
  &= \sum_{z^N}\sum_{w,s,t} \frac{1}{MKL_n}W_{Z|X}^{\otimes N}(z^N|x^N(w,s,t))\log \frac{W_{Z|X}^{\otimes N}(z^N|x^N(w,s,t)) }{Q_{Z^N}(z^N)}\nonumber\\
  &= \sum_{z^N}\sum_{w,s,t} \frac{1}{MKL_n}W_{Z|X}^{\otimes N}(z^N|x^N(w,s,t)) \log \frac{W_{Z|X}^{\otimes N}(z^N|x^N(w,s,t)) }{Q_{0}^{\otimes N}(z^N)}- \D{\widehat{Q}^N}{Q_0^{\otimes N}}\nonumber\\
  &\geq \sum_{z^N}\sum_{w,s,t} \frac{1}{MKL_n}W_{Z|X}^{\otimes N}(z^N|x^N(w,s,t)) \log \frac{W_{Z|X}^{\otimes N}(z^N|x^N(w,s,t)) }{Q_{0}^{\otimes N}(Z^N)} - \delta\nonumber\\
  &= \sum_{i=1}^N \sum_{z_i}\sum_{w,s,t}\frac{1}{MKL_n} W_{Z|X}(z_i|x_i(w,s,t)) \log \frac{W_{Z|X}(z_i|x_i(w,s,t)) }{Q_{0}(z_i)} - \delta\nonumber\\
  &\geq \sum_{i=1}^N \sum_{z_i}\sum_{w,s,t}\frac{1}{MKL_n} W_{Z|X}(z_i|x_i(w,s,t)) \log \frac{W_{Z|X}(z_i|x_i(w,s,t)) }{Q_{Z}(z_i)}+\sum_{i=1}^N\D{Q_{Z}}{Q_0}- \delta\nonumber\\
  &\geq \sum_{i=1}^N \sum_{z_i}\sum_{w,s,t}\frac{1}{MKL_n} W_{Z|X}(z_i|x_i(w,s,t))\log \frac{W_{Z|X}(z_i|x_i(w,s,t)) }{Q_{Z}(z_i)} - \delta\nonumber\\
              &= \sum_{i=1}^N \avgI{\widetilde{X}_i;\widetilde{Z}_i},
\end{align}
where $(\widetilde{X}_i,\widetilde{Z}_i)$ have joint distribution $P_{\widetilde{X}_i}(x)W_{Z|X}(z|x)$ induced by the marginal of the input distribution in the $i$-th position, i.e., $P_{\widetilde{X}_i}(x)=\sum_{w,s}\frac{\indic{x_i(w,s)=x}}{MK}$. Since conditioning does not increase entropy and $T-\widetilde{X}_i-\widetilde{Z}_i$, we have
\begin{align}
  \sum_{i=1}^N \avgI{\widetilde{X}_i;\widetilde{Z}_i} &\geq \sum_{i=1}^N \avgI{\widetilde{X}_i;\widetilde{Z}_i|T}\nonumber\\
                                                  &= \sum_{i=1}^N\sum_{t=1}^{L_n}\frac{1}{L_n}\avgI{\widetilde{X}_i;\widetilde{Z}_i|T=t}\nonumber\\
                                                  &\geq n\avgI{\overline{X};\overline{Z}},
\end{align}
where $(\overline{X},\overline{Z})$ have joint distribution $P_{X}(x)W_{Z|X}(z|x)$ induced by the marginal of the input distribution of the codewords within the slot used for transmission, i.e., $P_{X}(x)=\frac{1}{n}\sum_{i=1}^n\sum_{w,s}\frac{\indic{x^n(m,s)=x}}{MK}$ and $P_X(x_1)=\mu_n$. By~\cite[Lemma 1]{Bloch2016},
\begin{align*}
  n\avgI{\overline{X};\overline{Z}} \geq n\mu_n(\D{Q_1}{Q_0}-\epsilon)
\end{align*}
for $n$ large enough so that combining all the inequalities
\begin{align*}
  \lim_{n\to\infty}\frac{\log MK}{\sqrt{n\log L_n}} \geq \beta\sqrt{\frac{2}{\chi_2(Q_1\Vert Q_0)}}\D{Q_1}{Q_0}.
\end{align*}
Taking $\beta=\frac{1}{\sqrt{2}}$ yields the desired optimality on secret key bits when $r=\sqrt{\frac{1}{\chi_2(Q_1\Vert Q_0)}}\D{P_1}{P_0}$.
\newpage
\bibliographystyle{IEEEtran}
\bibliography{references.bib}

\begin{thebibliography}{10}
\providecommand{\url}[1]{#1}
\csname url@samestyle\endcsname
\providecommand{\newblock}{\relax}
\providecommand{\bibinfo}[2]{#2}
\providecommand{\BIBentrySTDinterwordspacing}{\spaceskip=0pt\relax}
\providecommand{\BIBentryALTinterwordstretchfactor}{4}
\providecommand{\BIBentryALTinterwordspacing}{\spaceskip=\fontdimen2\font plus
\BIBentryALTinterwordstretchfactor\fontdimen3\font minus
  \fontdimen4\font\relax}
\providecommand{\BIBforeignlanguage}[2]{{%
\expandafter\ifx\csname l@#1\endcsname\relax
\typeout{** WARNING: IEEEtran.bst: No hyphenation pattern has been}%
\typeout{** loaded for the language `#1'. Using the pattern for}%
\typeout{** the default language instead.}%
\else
\language=\csname l@#1\endcsname
\fi
#2}}
\providecommand{\BIBdecl}{\relax}
\BIBdecl

\bibitem{Arumugam2016a}
K.~S.~K. Arumugam and M.~R. Bloch, ``Keyless asynchronous covert
  communication,'' in \emph{Proc. of IEEE Information Theory Workshop},
  Cambridge, United Kingdom, Sep. 2016, pp. 191--195.

\bibitem{Bash2013}
B.~Bash, D.~Goeckel, and D.~Towsley, ``Limits of reliable communication with
  low probability of detection on {AWGN} channels,'' \emph{{IEEE} {J}ournal on
  {S}elected {A}reas in {C}ommunications}, vol.~31, no.~9, pp. 1921--1930, Sep.
  2013.

\bibitem{Wang2016b}
L.~Wang, G.~W. Wornell, and L.~Zheng, ``Fundamental limits of communication
  with low probability of detection,'' \emph{IEEE Transactions on Information
  Theory}, vol.~62, no.~6, pp. 3493--3503, Jun. 2016.

\bibitem{Bloch2016}
M.~R. Bloch, ``Covert communication over noisy channels: A resolvability
  perspective,'' \emph{IEEE Transactions on Information Theory}, vol.~62,
  no.~5, pp. 2334--2354, May 2016.

\bibitem{bouetteCovertCommunicationTwo2023}
C.~Bouette, L.~Luzzi, and L.~Wang, ``Covert {{Communication}} over {{Two
  Types}} of {{Additive Noise Channels}},'' in \emph{Proc. of {{IEEE
  Information Theory Workshop}}}, Saint-Malo, France, Apr. 2023, pp. 266--271.

\bibitem{Hou2014}
J.~Hou and G.~Kramer, ``Effective secrecy: {{Reliability}}, confusion and
  stealth,'' in \emph{Proc. of {{IEEE International Symposium}} on
  {{Information Theory}}}, Cambridge, Jun. 2014, pp. 601--605.

\bibitem{Lin2020Stealthy}
P.-H. Lin, C.~R. Janda, E.~A. Jorswieck, and R.~F. Schaefer, ``Stealthy secret
  key generation,'' \emph{Entropy}, vol.~22, no.~6, p. 679, jun 2020.

\bibitem{Lentner2020}
D.~Lentner and G.~Kramer, ``Stealth {{Communication}} with {{Vanishing Power}}
  over {{Binary Symmetric Channels}},'' in \emph{Proc. of {{IEEE International
  Symposium}} on {{Information Theory}}}, Jun. 2020, pp. 822--827.

\bibitem{Hou2017}
J.~Hou, G.~Kramer, and M.~R. Bloch, ``Effective {{Secrecy}}: {{Reliability}},
  {{Confusion}}, and {{Stealth}},'' in \emph{Information {{Theoretic Security}}
  and {{Privacy}} of {{Information Systems}}}, R.~F. Schaefer, H.~Boche,
  A.~Khisti, and H.~V. Poor, Eds.\hskip 1em plus 0.5em minus 0.4em\relax
  Cambridge University Press, 2017.

\bibitem{Wang2021}
S.-Y. Wang and M.~R. Bloch, ``Covert mimo communications under variational
  distance constraint,'' \emph{IEEE Transactions on Information Forensics and
  Security}, vol.~16, pp. 4605 -- 4620, Sep. 2021.

\bibitem{Bendary2021}
A.~Bendary, A.~Abdelaziz, and C.~E. Koksal, ``Achieving positive covert
  capacity over {MIMO} {AWGN} channels,'' \emph{{IEEE} Journal on Selected
  Areas in Information Theory}, vol.~2, no.~1, pp. 149--162, Mar. 2021.

\bibitem{Gagatsos2020Covert}
C.~N. Gagatsos, M.~S. Bullock, and B.~A. Bash, ``Covert capacity of bosonic
  channels,'' \emph{IEEE Journal on Selected Areas in Information Theory},
  vol.~1, no.~2, pp. 555--567, Aug. 2020.

\bibitem{Zlotnick2023Entanglement}
E.~Zlotnick, B.~Bash, and U.~Pereg, ``Entanglement-assisted covert
  communication via qubit depolarizing channels,'' in \emph{Proc. of IEEE
  International Symposium on Information Theory}, Taipei, Taiwan, Jun. 2023,
  pp. 198--203.

\bibitem{Wang2024Resource}
S.-Y. Wang, S.-J. Su, and M.~R. Bloch, ``Resource-efficient
  entanglement-assisted covert communications over bosonic channels,'' in
  \emph{Proc. of IEEE International Symposium on Information Theory}, Athens,
  Greece, Jul. 2024, pp. 3106--3111.

\bibitem{Sobers2017}
T.~V. Sobers, B.~A. Bash, S.~Guha, D.~Towsley, and D.~Goeckel, ``Covert
  {{Communication}} in the {{Presence}} of an {{Uninformed Jammer}},''
  \emph{IEEE Transactions on Wireless Communications}, vol.~16, no.~9, pp.
  6193--6206, Sep. 2017.

\bibitem{Lee2018a}
S.~H. Lee, L.~Wang, A.~Khisti, and G.~W. Wornell, ``Covert communication with
  channel-state information at the transmitter,'' \emph{IEEE Transactions on
  Information Forensics and Security}, vol.~13, no.~9, pp. 2310--2319, Sep.
  2018.

\bibitem{ZivariFard2020}
H.~Zivari-Fard, M.~Bloch, and A.~Nosratinia, ``Keyless covert communication via
  channel state information,'' \emph{IEEE Transactions on Information Theory},
  vol.~68, no.~8, pp. 5440--5474, Aug. 2022.

\bibitem{Che2014a}
P.~H. Che, M.~Bakshi, C.~Chan, and S.~Jaggi, ``Reliable deniable communication
  with channel uncertainty,'' in \emph{Proc. of IEEE Information Theory
  Workshop}, Hobart, Tasmania, Nov. 2014, pp. 30--34.

\bibitem{Hayashi2024Covert}
M.~Hayashi and A.~V\'azquez-Castro, ``Covert communication with gaussian noise:
  From random access channel to point-to-point channel,'' \emph{IEEE
  Transactions on Communications}, vol.~72, no.~3, pp. 1516--1531, Mar. 2024.

\bibitem{Tchamkerten2009}
A.~Tchamkerten, V.~Chandar, and G.~W. Wornell, ``Communication under strong
  asynchronism,'' \emph{IEEE Transactions on Information Theory}, vol.~55,
  no.~10, pp. 4508--4528, Oct. 2009.

\bibitem{Bash2016}
B.~A. Bash, D.~Goeckel, and D.~Towsley, ``Covert communication gains from
  adversary's ignorance of transmission time,'' \emph{IEEE Transactions on
  Wireless Communications}, vol.~15, no.~12, pp. 8394--8405, Dec. 2016.

\bibitem{Goeckel2016}
D.~Goeckel, B.~Bash, S.~Guha, and D.~Towsley, ``Covert communications when the
  warden does not know the background noise power,'' \emph{IEEE Communications
  Letters}, vol.~20, no.~2, pp. 236--239, Feb. 2016.

\bibitem{Dani2021Covert}
V.~Dani, V.~Ramaiyan, and D.~Jalihal, ``Covert communication over asynchronous
  channels with timing advantage,'' in \emph{Proc. of IEEE Information Theory
  Workshop}, India, Oct. 2021.

\bibitem{ibrahimIdentificationEffectiveSecrecy2021}
A.~Ibrahim, R.~Ferrara, and C.~Deppe, ``Identification under {{Effective
  Secrecy}},'' in \emph{Proc. of {{IEEE Information Theory Workshop}}}, Oct.
  2021, pp. 1--6.

\bibitem{rosenbergerCapacityBoundsIdentification2023}
J.~Rosenberger, A.~Ibrahim, B.~A. Bash, C.~Deppe, R.~Ferrara, and U.~Pereg,
  ``Capacity {{Bounds}} for {{Identification With Effective Secrecy}},'' in
  \emph{Proc. of {{IEEE International Symposium}} on {{Information Theory}}},
  Taipei, Taiwan, Aug. 2023.

\bibitem{Tahmasbi2019}
M.~Tahmasbi and M.~R. Bloch, ``First and second order asymptotics in covert
  communication,'' \emph{IEEE Transactions on Information Theory}, vol.~65,
  no.~4, pp. 2190 --2212, Apr. 2019.

\bibitem{lehmann2006testing}
E.~L. Lehmann and J.~P. Romano, \emph{Testing {{Statistical
  Hypotheses}}}.\hskip 1em plus 0.5em minus 0.4em\relax Springer New York,
  2006.

\bibitem{Zhang2019a}
Q.~E. Zhang, M.~R. Bloch, M.~Bakshi, and S.~Jaggi, ``Undetectable {{Radios}}:
  {{Covert Communication}} under {{Spectral Mask Constraints}},'' in
  \emph{Proc. of {{IEEE International Symposium}} on {{Information Theory}}},
  Paris, France, Jul. 2019, pp. 992--996.

\bibitem{Han1993}
T.~Han and S.~Verdu, ``Approximation theory of output statistics,'' \emph{IEEE
  Transactions on Information Theory}, vol.~39, no.~3, pp. 752--772, May 1993.

\bibitem{Cuff2013}
P.~Cuff, ``Distributed {{Channel Synthesis}},'' \emph{IEEE Transactions on
  Information Theory}, vol.~59, no.~11, pp. 7071--7096, Nov. 2013.

\bibitem{boucheronConcentrationInequalitiesNonasymptotic2013}
S.~Boucheron, G.~Lugosi, and P.~Massart, \emph{Concentration {{Inequalities}}:
  {{A Nonasymptotic Theory}} of {{Independence}}}.\hskip 1em plus 0.5em minus
  0.4em\relax Oxford University Press, Feb. 2013.

\bibitem{Chandar2013}
V.~Chandar, A.~Tchamkerten, and D.~Tse, ``Asynchronous {{Capacity}} per {{Unit
  Cost}},'' \emph{IEEE Transactions on Information Theory}, vol.~59, no.~3, pp.
  1213--1226, Mar. 2013.

\bibitem{Watanabe2014Strong}
S.~Watanabe and M.~Hayashi, ``Strong converse and second-order asymptotics of
  channel resolvability,'' in \emph{Proc. of IEEE International Symposium on
  Information Theory}, Jun. 2014, pp. 1882--1886.

\bibitem{Polyanskiy2010a}
Y.~Polyanskiy, H.~V. Poor, and S.~Verdu, ``Channel {{Coding Rate}} in the
  {{Finite Blocklength Regime}},'' \emph{IEEE Transactions on Information
  Theory}, vol.~56, no.~5, pp. 2307--2359, May 2010.

\bibitem{Arumugam2019}
K.~S.~K. Arumugam and M.~R. Bloch, ``Covert {{Communication Over}} a {{K}}
  -{{User Multiple-Access Channel}},'' \emph{IEEE Transactions on Information
  Theory}, vol.~65, no.~11, pp. 7020--7044, Nov. 2019.

\bibitem{wangGaussianCovertCommunication2019}
L.~Wang, ``On {{Gaussian}} covert communication in continuous time,''
  \emph{EURASIP Journal on Wireless Communications and Networking}, vol. 2019,
  no.~1, p. 283, Dec. 2019.

\bibitem{Sason2016}
I.~Sason and S.~Verdu, ``{$f$} -{{Divergence Inequalities}},'' \emph{IEEE
  Transactions on Information Theory}, vol.~62, no.~11, pp. 5973--6006, Nov.
  2016.

\bibitem{Bounhar2024Whispering}
A.~Bounhar, M.~Sarkiss, and M.~Wigger, ``Whispering {{Secrets}} in a {{Crowd}}:
  {{Leveraging Non-Covert Users}} for {{Covert Communications}},'' \emph{arXiv
  preprint}, vol. 2408.12962, Aug. 2024.

\end{thebibliography}

\end{document}